%% file: main_sp.tex
\documentclass[conference,compsoc]{IEEEtran}
\ifCLASSOPTIONcompsoc
  \usepackage[nocompress]{cite}
\else
  \usepackage{cite}
\fi

\usepackage[utf8]{inputenc} 
\usepackage[T1]{fontenc}    
\usepackage{hyperref}       
\usepackage{url}            
\usepackage{booktabs}       
\usepackage{amsfonts}       
\usepackage{nicefrac}       
\usepackage{microtype}      
\usepackage{xcolor}         

\input{macros.tex}
\begin{document}

\title{More is Merrier: Relax the Non-Collusion Assumption in Multi-Server PIR}

\author{
\IEEEauthorblockN{Tiantian Gong\IEEEauthorrefmark{1}, Ryan Henry\IEEEauthorrefmark{2}, Alexandros Psomas\IEEEauthorrefmark{1}, Aniket Kate\IEEEauthorrefmark{1}\IEEEauthorrefmark{3}}
\IEEEauthorblockA{\IEEEauthorrefmark{1}Purdue University \emph{tiangong17@gmail.com, \{apsomas, aniket\}@purdue.edu}}
\IEEEauthorblockA{\IEEEauthorrefmark{2}University of Calgary \emph{ryan.henry@ucalgary.ca}, \IEEEauthorrefmark{3}Supra Research}
}

\maketitle

\begin{abstract}
\input{sections/1_abstract}
\end{abstract}

\IEEEpeerreviewmaketitle

\pagestyle{plain}

\input{sections/2_intro-sp}
\input{sections/3_overview}
\input{sections/5_collusion-k-servers}

\input{sections/7_privacy}
\input{sections/6_implementation}
\input{sections/9_literature}
\input{sections/10_conclusion}
\input{sections/12_ack}

\bibliographystyle{IEEEtranS}
\bibliography{all}

\appendices
\input{sections/appendix-dishonest.tex}

\input{sections/appendix-constructions.tex}
\input{sections/appendix-selfinsurance.tex}


\end{document}

%% file: macros.tex
\usepackage[utf8]{inputenc}

\usepackage{comment}
\usepackage{algpseudocode}
\usepackage{bbm}

\usepackage{amsmath, amsthm, amstext, comment, graphicx}
\usepackage{mathtools,extarrows}
\usepackage{subfiles,paralist}
\usepackage{tikz}
\usepackage{pgfplots}
\pgfplotsset{compat=1.15}
\usepackage{mathrsfs}
\usepackage{mdframed}
\usetikzlibrary{arrows}

\usepackage[ruled,linesnumbered,noend]{algorithm2e} 
\usepackage{url}

\allowdisplaybreaks
\usepackage{hyperref}
\hypersetup{colorlinks=true,linkcolor=black,citecolor=blue}
\usepackage{enumitem}

\usepackage{thmtools, thm-restate}
\usepackage{float}
\usepackage{caption}
\usepackage{subcaption}

\usepackage{multirow}
\usepackage[ruled,linesnumbered]{algorithm2e} 
\usepackage[most]{tcolorbox}
\usepackage{adjustbox}
\usepackage{cleveref}
\usepackage{csquotes}
\usepackage{wrapfig}
\usepackage{soul}

\SetAlFnt{\small}
\SetAlCapFnt{\small}
\SetAlCapNameFnt{\small}
\SetAlCapHSkip{0pt}
\IncMargin{-\parindent}

\newtheorem{definition}{Definition}

\newtheorem{corollary}{Corollary}

\renewcommand\paragraph[1]{\noindent \textbf{#1.\ }}

\renewcommand{\vec}[1]{\mathbf{#1}}

\newcommand*{\rom}[1]{\expandafter\@slowromancap\romannumeral #1@}
\newcommand{\RNum}[1]{\uppercase\expandafter{\romannumeral #1\relax}}

\newcommand{\tian}[1]{{\color{black}#1}}

\newcommand{\circno}[1]{\textcircled{\raisebox{-0.9pt}{#1}}}
\newcommand{\D}{\mathbb{D}} 
\newcommand{\fake}{companion }

\newcommand{\amt}[1]{{\color{blue}$\lambda_{#1}$}}
\newcommand{\amtt}[1]{{\color{blue}\lambda_{#1}}}

\newcommand{\vcgmath}{V }

\newcommand{\aarnr}{\vcgmath + \amtt{s} + \amtt{r}}

\newcommand{\client}{{\color{purple}{$\mathsf{U}$}}}

\usepackage{enumitem}
\setlist{topsep=0pt}
\RequirePackage[normalem]{ulem} 
\RequirePackage{color}\definecolor{RED}{rgb}{1,0,0}\definecolor{BLUE}{rgb}{0,0,1} 

%% file: sections/1_abstract.tex
A long line of research on secure computation has confirmed that anything that can be computed, can be computed securely using a set of non-colluding parties.
Indeed, this non-collusion assumption makes a number of problems solvable, as well as reduces overheads and bypasses computational hardness results, and it is pervasive across different privacy-enhancing technologies. However, it remains highly susceptible to covert, undetectable collusion among computing parties. This work stems from an observation that if the number of available computing parties is much higher than the number of parties required to perform a secure computation task, collusion attempts in privacy-preserving computations could be deterred. 

We focus on the prominent privacy-preserving computation task of multi-server $1$-private information retrieval (PIR) that inherently assumes no pair-wise collusion. For PIR application scenarios, such as those for blockchain light clients, where the available servers can be plentiful, a single server's deviating action is not tremendously beneficial to itself. We can make deviations undesired via small amounts of rewards and penalties, thus significantly {\em raising the bar} for collusion resistance. We design and implement a collusion mitigation mechanism on a public bulletin board with payment execution functions, considering only rational and malicious parties with no honest non-colluding servers. Privacy protection is offered for an extended period after the query executions. 

%% file: sections/2_intro-sp.tex
\section{Introduction}\label{sec: pir intro}
The non-collusion assumption among computing parties is prevalent across privacy-preserving computations. 
It allows us to solve some important problems~\cite{cleve1986limits,TwoSources,LamportSP82}, with no known solutions without it. It also reduces overheads and allows us to bypass computational hardness results in other problems such as private information retrieval (PIR)~\cite{chor1995private}. In fact, it is fair to say that, the entire field of secure multi-party computation (MPC)~\cite{Yaosecuremultipartycomputation,BGW88} relies on the non-collusion assumption. 
While most of these distributed systems do allow collusion among pre-defined subsets of parties, all bets are off once the adversary can lure a superset of parties. 
Indeed, using the non-collusion assumption remains highly susceptible to undetectable collusion among computing parties: collusion over Internet-based secure systems can be difficult to prove, if possible; parties may collude without recognizing each other. 
As these distributed systems get considered for applications such as privacy-preserving machine learning~\cite{SecureML,GAZELLE,wagh2019securenn,byali2020flash,knott2021crypten,chaudharitrident}, privacy-preserving database/directory searches~\cite{6956575,MittalOTBG11}, and blockchain privacy~\cite{henry2018blockchain,qin2019applying}, the non-collusion assumption is considered an Achilles' heel for large adoption.

In the context of database querying service, a PIR protocol allows a client to query a database without revealing which piece of information is of interest. Chor et al.\ \cite{chor1995private} formalize the $k$-server PIR problem ($k\geq 2$) and propose a mechanism for fetching a single bit from an $N$-bit string, which is stored by $\ell \geq k$ servers (or parties/players), without disclosing the bit's location to the $k$ queried servers.

PIR can be constructed from a single server or multiple servers. Single server information-theoretic PIR (IT-PIR) is impossible beyond the trivial solution of sending the whole database to the user. The \textit{inefficiency} of single-server computational PIR (cPIR) \cite{kushilevitz1997replication} schemes is also well-known. 
Consider a database $\D$ comprising $N$ entries. 
The state-of-the-art proposals~\cite{angel2018pir,ahmad2021addra,corrigan2022single,henzinger2022one,menon2022spiral} usually have $O(\sqrt{N})$ communication complexity. OnionPIR~\cite{mughees2021onionpir} realizes $O(\log N)$ communication cost while incurring heavy computation costs. (Detailed efficiency comparisons reside in Appendix \ref{sec:constructions}.) 
By involving more \textit{non-colluding} servers ($k\geq 2$), the communication and computation complexity can be substantially reduced, which serves as a motivation for multi-server PIR schemes~\cite{chee2013query,shah2014one,boyle2016function,hafiz2019bit}.
For example, in Hafiz-Henry's multi-server PIR~\cite{hafiz2019bit}, the query size is $130\log k (\lceil\log N \rceil -7)$ bits and the total response size has only $\frac{ k}{k-1}$ times overhead. It also has the lowest possible computation cost~\cite{beimel2000reducing}: for each bit being fetched, each server performs $\frac{|\D|}{k-1}$ bit operations on the database in expectation. 

However, the non-collusion assumption promoting efficiency improvements calls for practical justification. 
Because servers can easily collude over unobserved side channels, or various covert and anonymous channels~\cite{syverson2004tor} to recover users' queried entries, and such \textbf{privacy-targeted collusion} remains oblivious to PIR procedures. The prevalence of these communication modalities makes the collusion problem impossible to solve within the PIR protocol. 

In this paper, we find that by relaxing the non-collusion assumption to the rationality assumption on servers, deterring such privacy-targeted collusion becomes possible. Here, a rational party takes actions that maximize its utility. 
The core question is then \textit{how to render collusion undesired for rational participants}, for which we draw intuition from a novel bit guessing game.

\paragraph{Secret-shared bit guessing game} 
An honest game host commits to a secret unbiased bit $b$ and secret-shares it among $k$ players. Here, every player receives a share such that bit $b$ can be reconstructed with all $k$ shares together. After sharing, each player can make \textit{open} guesses about $b$ and the game terminates when no more player wishes to guess. 
The \textit{first} correct guesser is selected as the winner. The host rewards the winner \$2 but takes \$4 from the other $k-1$ players. 
Additionally, a player loses \$4 for a wrong guess. Without reconstructing bit $b$ explicitly from received shares, rational players do not guess because the expected return is negative (\$-1). 
For players without a network advantage in submitting guesses first, not collaborating to reconstruct $b$ is strictly dominating (i.e., generating strictly more returns) since their expected return is $(\frac{k-1}{k}(-4)+\frac{1}{k}\cdot 2)$ which is negative for $k\geq 2$. This thought experiment implies that it is possible to make unobserved collusion visible and undesired for rational participants via a proper incentive structure.

\paragraph{Deter collusion in multi-server PIR}
The simple game above is already expressive, but there are caveats that can cause the incentives to stop being effective in practice, e.g., a rational host or bit $b$ being biased. When designing a collusion-mitigation mechanism for multi-server PIR, we then face two challenges: (i) the difficulty of defining and verifying evidence of collusion when users and servers are not honest, and (ii) the intricate conflicts between ensuring the effectiveness of collusion mitigation and deterring manipulations from users and servers for biased and valuable secrets. Instead of making assumptions on collusion methods, we base the solution only on the premise that \textit{at least some colluding parties learn user secrets or some nontrivial (i.e., output depends on the inputs) function of the secrets after successful collusion}. 
We then define collusion evidence to be outputs of non-trivial functions on user secrets and record commitments (e.g., cryptographic hash) of queries and responses for evidence verification.

For the incentives, users pay service fee $\amtt{s}$ for servers' PIR responses. We reward the colluding party (with $\amtt{r}$) who reports collusion first and penalize the remaining accomplices by exerting penalty $\amtt{p}$ and depriving them of service fees. To avoid manipulations of the mechanism, we add randomness and apply fines $\amtt{f}$ to discourage servers with private knowledge about user secrets from making false accusations. We tune $\amtt{s}$ (when necessary) to hinder users' fake accusations. For parameterization of the payment amounts, we note the following:

\underline{In a single or known finite runs of the PIR service}, we do not need many servers, i.e., $\ell = k$, or reporting rewards, i.e., $\amtt{r}= 0$. But in the worst case, we need to calibrate the service fee and penalty amounts such that their weighted sum, $\amtt{s} + \frac{k-1}{k} \amtt{p}$, is greater than the secret's worth. This means that one expects to earn more by not colluding and earning the service fee. A larger $k$ means looser condition on \amt{s} and \amt{p}, expanding parameter feasibility regions. 
Since $\amtt{r}= 0$, servers and users do not falsely accuse when there are positive fines $\amtt{f} > 0 $ and service fees $\amtt{s} > 0$. 

\underline{In infinitely repeated runs of the service}, we need to have many servers, i.e., $\ell \gg k$, and sufficiently positive reporting rewards, i.e., $\amtt{r} > X$ where $X$ is a positive term that decreases with $\ell$ (\Cref{thm: k agent l server repeated}). Intuitively, to achieve the non-collusion outcome in repeated runs, the same servers need to be queried less frequently to reduce the privacy worth they bear. And servers need to earn positive premiums by deviating from an undesired strategy emerging in repeated games, which is always cooperating and not reporting. 
Since $\amtt{r} > 0$, the fines $\amtt{f}$ and service fees $\amtt{s}$ now both need to exceed the reward by a certain amount to prevent false accusations.

\subsection{Contributions}
We devise a mechanism for relaxing the non-collusion assumption in multi-server PIR systems to rationality assumptions on servers. It is composed of \textit{a practical incentive structure} and \textit{a module for gathering and verifying collusion evidence}. 
Overall, we make the following contributions. 

First and most importantly, we present a \textbf{practical collusion deterrence mechanism for multi-server PIR} so that one can enjoy the perfect efficiency of 1-private PIR.\footnote{1-private PIR schemes are constructed assuming no pairwise collusion and provide effectively perfect efficiency in terms of downloading capacity, in the sense that to fetch a $b$-bit entry, one only needs to download $b+1$ bits~\cite{shah2014one} (by having exponentially many servers). Hafiz-Henry's construction~\cite{hafiz2019bit} of 1-private multi-server PIR achieves the best possible download cost, $\frac{k}{k-1} N b $ bits, with a fixed number of servers.} 
(The mechanism directly applies to general $t$-private PIR that only assumes non-collusion among $>t$ parties.) 
This is achieved in harsh settings: (1) We let there be a secure strong collusion protocol at colluding parties' disposal. They can compute arbitrary functions on received queries or computed responses, e.g., reconstructing the queried entry. (2) Servers can hold \textit{arbitrary} private knowledge about clients' queries. (3) Servers can collude over \textit{any unobserved channels}. (4) A bounded number of servers are malicious. 
We employ many servers so that a single server does not bear too much information worth leaking before it is discouraged to collude (similar to when the share holders hesitate to recover bit $b$ in the guessing game). 

Second, we examine how servers exit the PIR service and adopt a privacy protection self-insurance to \textbf{protect user secrets for an extended period} after the queries. 
Third, we study how servers form strong coalitions, whose members have complete trust for each other and are immune to the provided incentives, from a cooperative game theory perspective. 
Our focus is identifying the conditions necessary to keep the coalition size small. 
Last, we \textbf{implement and deploy the mechanism} as a smart contract on Ethereum~\cite{yellow} blockchain and apply optimizations to economize the gas costs (in \Cref{tab:gas}).

\tian{
\paragraph{General applicability to additive secret sharing style applications}
We note that our deterrence mechanism and game-theoretic analysis apply to general additive secret sharing style applications, i.e., all parties need to provide input to learn non-trivial information about the secret.\footnote{The mechanism also applies to general access structures where subsets of parties can learn non-trivial information about the secret, but non-colluding parties could be negatively affected if the number of malicious parties exceeds our tolerance.} 
Specifically, as hinted in the secret-shared bit guessing game and shown in \Cref{sec:realtime}, our definition of the collusion game and the game-theoretic analysis do not depend on details of the underlying PIR protocol. 
}

\paragraph{Efficiency and practicality}
Overall, the solution only adds to the underlying PIR scheme one signed commitment message for each query and response. The collusion resolution mechanism also has little overhead: it induces the non-collusion outcome among the rational servers and is not executed in equilibrium when there is no malicious server. 
For practicality, we observe that in many settings where the number of available computing parties can be much higher than the required set, an effective collusion mitigation mechanism with practical parameters, i.e., small fee \amt{s} and penalty \amt{p}, can be constructed.

\paragraph{Relevance to Blockchain Privacy}
As a representative example, the proposed mechanism is useful for blockchain access privacy problem~\cite{henry2018blockchain}: allowing light/thin clients that do not maintain the ledger locally to query full nodes in a privacy-preserving way~\cite{qin2019applying}. 
Existing solutions are either inefficient given the ever-changing nature of database~\cite{qin2019applying} or require trusted hardware~\cite{le2020tale,matetic2019bite,wust2019zlite}. 
Blockchains are particularly suitable application scenarios for our mechanism as prominent blockchains have thousands of full nodes~\cite{bitnode,ethnode}, which admits more practical parameters for the design.

%% file: sections/3_overview.tex
\section{Preliminaries and overview}\label{sec: prelim}
\subsection{Definitions}
\paragraph{Multi-server PIR}
We follow the security definition for multi-server PIR by Hafiz-Henry~\cite{hafiz2019bit}. Let $I,Q,R,Y$ respectively denote the random variables for the entry index $i\in [N]$ of interest, the query string $q\in \{0,1\}^*$, the response string $r\in \{0,1\}^*$ and the reconstructed entry $y\in \D$. 

\begin{definition}[Multi-server PIR~\cite{hafiz2019bit}]\label{def:pir}
The interaction described by $(I,Q,R,Y)$ provides correctness if $\Pr[Y=\D_i|I=i]=1$ where $\D_i$ is the $i$-th entry in $\D$. 
It provides perfect 1-privacy if $\forall i, i^*\in [N], \Pr[Q=q|I=i]=\Pr[Q=q|I=i^*]$ where the probability is over all the random coin tosses made by the client. 
It provides computational 1-privacy if $\forall i, i^*\in [N]$, the distribution ensembles $\{Q|I=i\}_{\lambda \in \mathbb{N}}$ and $\{Q|I=i^*\}_{\lambda \in \mathbb{N}}$ are computationally indistinguishable with $\lambda$ being the security parameter.
\end{definition}

We focus on $k$-out-of-$k$ 1-private PIR schemes where all $k$ responses are needed in reconstruction. Note that the analysis applies to $k'$-out-of-$k$ ($k'<k$) robust PIR schemes, but non-colluding parties can be penalized in the worst case (but not in equilibrium). We then denote the multi-server PIR scheme as $(\ell,k,1)$-PIR. A set of $\ell$ servers $S=\{S_i | i\in [\ell] \}$ maintain identical databases consisting of $N$ entries, $\D=\{\D_i | i\in [N]\}$. To query entry $\D_a$ at index $a$, user \client \ generates $k$ queries 
and sends them to $k$ (distinct) servers. Each queried server locally computes responses to the query and responds with an answer string. \client \ then collects all $k$ response strings and reconstructs $\D_a$.

There is an abundance of available PIR constructions, including the ones for two servers \cite{chor1995private,wehner2005improved,dvir20162,razborov2006omega,boyle2016function}, and three or more servers \cite{ambainis1997upper,beimel2002breaking,yekhanin2008towards,efremenko20123,itoh2010improved,tajeddine2018private,kumar2017private,chee2013query,hafiz2019bit}. While all constructions suffice for this work, we find Boyle-Gilboa-Ishai's construction for 2-server cPIR~\cite{boyle2016function} and Hafiz and Henry's construction for more server case~\cite{hafiz2019bit} to be suitable. 
Both constructions realize PIR as a distributed point function and as elaborated in Appendix \ref{sec:constructions}, are highly efficient for million-sized databases.

\paragraph{Relevant concepts in game theory~\cite{myerson1997game}}
A normal-form game can be characterized with a set of players $\mathcal{P}$, actions available to players $\mathcal{A}=\times_{i\in \mathcal{P} } \mathcal{A}_{i}$, and utility functions $\mathcal{V}=\{u_1, \dotsi, u_{|\mathcal{P}|}\}$ where $u_i: \mathcal{A} \mapsto \mathbb{R}$, mapping action profiles to real-numbered utilities. An extensive-form game further captures sequences of actions, beliefs over others' actions, knowledge of history, etc. 

Let $\vec{s}\in \mathcal{A}$ be a strategy profile, which is simply a sample in universe $\mathcal{A}$. A \textit{dominant strategy} $\vec{s}_i$ for a player $\mathcal{P}_i$ generates utility at least as good as any other strategies: $\forall \vec{s}_i' \neq \vec{s}_i, \forall \vec{s}_{-i}, u_i(\vec{s}_i, \vec{s}_{-i}) \geq u_i(\vec{s}_i', \vec{s}_{-i}) $ where $\vec{s}_{-i}$ is the strategy profile for all other players. If equality never holds, then $\vec{s}_i$ is a strictly dominant strategy. 
A \textit{solution concept} then captures the equilibrium strategies for all players in a game. 
Our incentive structure induces a (extensive-form) sequential game and the solution concept we employ is subgame perfect equilibrium and sequential equilibrium, which we discuss more in \Cref{sec:realtime} and \Cref{sec:deceitful}. Informally, in both equilibria, at each decision point of the sequential game, a player selects the action that maximizes the expected returns from that point on. 

\begin{figure*}[ht]
    \centering
    \begin{mdframed}
    \centering
    \textbf{Normal Service (for querying $\D_a$)}
    \begin{flushleft}{
    \begin{enumerate}[wide, labelindent=0pt]
    \item User \client \ samples an index $a_r$ uniformly at random and runs two instances of $(\ell, k, 1)$-PIR for querying $\D_a$ and $\D_{a_r}$, generating $2k$ queries, $Q_1,\ldots,Q_k$ for $\D_a$ and $Q'_1,\ldots,Q'_k$ for $\D_{a_r}$. 
    \item \client \ commits to the $2k$ queries, obtaining $\vec{c} = (\langle c_1,\sigma_1\rangle ,\ldots, \langle c_k, \sigma_k\rangle)$ and $\vec{c'} = (\langle c'_1, \sigma'_1\rangle,\ldots, \langle c'_k, \sigma'_k\rangle)$. 
    \client \ picks a random permutation $\Pi$, posts $\Pi(\vec{c},\vec{c'})$ to \textbf{BB}, and stores $\Pi$ locally. Here, for $i \in [k]$, $c_i = \mathsf{Comm}(Q_i)$, $c'_i = \mathsf{Comm}( Q'_i)$, $\sigma_i = \mathsf{Sign}( c_i)$ and $\sigma'_i = \mathsf{Sign}( c'_i)$. 
    \item \client \ samples $k$ distinct integers in $[\ell]$ at random and obtains $k$ corresponding server indices, $\vec{id} = (id_1,\ldots, id_{k})$. \client \ signs and posts $\langle \vec{id}, \sigma_{id}\rangle$ onto \textbf{BB}. Here, $\sigma_{id} = \mathsf{Sign}(\vec{id})$. 
    \item \client \ sends corresponding de-commit information to each server $S_{id_i}$ over a secure channel. 
    \item Each queried server $S_{id_i}$ verifies and de-commits the $i$-th commitments, and computes the answers $(A_i^1, A_i^2)$ to de-committed queries locally against database $\D$. 
    \item Each queried server $S_{id_i}$ commits to the responses, signs the message and posts $(\langle c_{id_i}^1, \sigma_{id_i}^1\rangle, \langle c_{id_i}^2, \sigma_{id_i}^2\rangle)$ onto \textbf{BB}. $S_{id_i}$ sends de-commit information to \client \ over a secure channel. Here, $c_{id_i}^1 = \mathsf{Comm}(A_i^1)$, $c_{id_i}^2 = \mathsf{Comm}(A_i^2)$, $\sigma_{id_i}^1 = \mathsf{Sign}(c_{id_i}^1)$ and $\sigma_{id_i}^2 = \mathsf{Sign}(c_{id_i}^2)$. 
    \item \client \ verifies and de-commits the commitments, retrieves responses, and reconstructs entry $\D_a$.
    \end{enumerate}}
    \end{flushleft}
    \smallskip
    
    \textbf{Collusion Resolution (Server $S_{id_i}$ accusing)}
    \begin{flushleft}{
    \begin{enumerate}[wide, labelindent=0pt]
    \item[8)] $S_{id_i}$ submits a collusion report to \textbf{CC}, indicating the evidence $\mathsf{ed}$, i.e., circuits for computing $f(\cdot)$, $S_{id_i}$'s inputs (i.e., a query or answer string) to compute $f(\cdot)$, and function outputs $\mathsf{ed}$. \textbf{CC} rejects the report if $f(\cdot)$ is trivial. 
    \item[9)] Each accused server $S_{id_j}$ provides auxiliary de-commit information for all possible input(s) to $f(\cdot)$. 
    \item[10)] \textbf{CC} confirms the accusation if (1) the accused server(s) fails to submit auxiliary information in time $\Delta$, or (2) the de-commit info is incorrect, or the output of $f(\cdot)$ on de-committed inputs match $\mathsf{ed}$. 
    \end{enumerate}}
    \end{flushleft}
    \end{mdframed}
    \caption{Overview of routines ($\omega = 1$ \fake query) with payments omitted. 
    $\mathsf{Comm}(\cdot)$ is a perfectly hiding and computationally binding 
    commitment scheme and $\mathsf{Sign}(\cdot)$ is a secure digital signature scheme.}
    \label{fig:solution}
\end{figure*}

\subsection{Threat model} 
We assume all servers are either rational, i.e., taking actions that maximize utilities, or malicious, i.e., behaving arbitrarily. We assume static corruption when faced with a malicious adversary. 
There are {\em no} honest non-colluding servers. 
Normally, servers provide PIR services by running a secure PIR protocol and are compensated by the system in the form of service fees. They can potentially collude with each other to break user privacy and obtain what the recovered private information is worth. Servers can collude as individuals (\Cref{sec:realtime} and \ref{sec: malicious}), form mutually-trusted coalitions (\Cref{sec: coalition}), or leave the service (\Cref{sec: exiting}).

To make the results more general, we {\em handicap} ourselves by giving the adversary advantages in collusion. If these advantageous settings are not achieved in practice, the result still holds. First, servers may hold arbitrary knowledge about a user's index of interest. 
Second, servers can communicate with each other over any unobserved communication channels, e.g., anonymous channels that keep the communication covert and undetectable. 
Further, we assume the communication is simulatable so that a server can generate the communication transcripts on its own, and colluding servers can self-protect by denying having the conversation. Third, we let there exist a secure collusion protocol, e.g., a secure MPC protocol, that can be utilized by colluding parties to compute the result collectively.

\subsection{Solution overview}\label{sec:overview}
We describe the solution given rational servers and present pseudocodes in \Cref{fig:solution}. Servers are deterministically indexed and participants all know the indices. We make use of a public bulletin board (\textbf{BB}) with payment execution functionality to instantiate the mechanism as a coordinator contract (\textbf{CC}). 
Algorithm \textbf{CC} verifies collusion evidence and executes payments involved in both normal service and collusion resolution. One implementation of \textbf{CC} is a smart contract on Ethereum~\cite{yellow} (\Cref{sec: notes on implementation}).

\paragraph{Setup and normal service}
Users know servers' identities, and they can directly communicate. 
To mitigate collusion, we have participants post commitments of messages on \textbf{BB} for potential evidence verification. Since we allow servers to hold arbitrary private knowledge, $\omega\geq 1$ random companion queries are sent to help distinguish between servers learning user secrets via collusion and via private knowledge. 

We now describe the normal routines with one random companion query. 
Initially, as described in line 1 in \Cref{fig:solution}, a user samples one random index $a_r$ and generates queries for both the intended entry $\D_a$ and the random entry $\D_{a_r}$. 
The user then commits to the $k$ pairs of queries, randomly permutes each pair, and posts the commitments to \textbf{BB} (line 2). Next, the user samples $k$ servers randomly and send corresponding de-commit information to them (line 3-4). 
Queried servers then compute the responses (line 5), post commitments of responses to \textbf{BB} and send de-commit information to the user over any secure channels (line 6). The user then reconstructs $\D_a$ from the responses (line 7).

\paragraph{Collusion} 
Colluding servers can collectively execute any collusion protocol to learn some nontrivial function $f(\cdot)$ of the explicit bits of the index or entry (e.g., least significant bit) or learn other servers' de-committed queries or responses. 

\paragraph{Collusion resolution} 
A collusion reporter provides the following as evidence: the circuit for computing function $f(\cdot)$, her inputs to $f(\cdot)$, and the output from computing $f(\cdot)$ (line 8 in \Cref{fig:solution}). 
Upon receiving a report, the contract confirms the non-triviality of $f(\cdot)$ (details in \Cref{sec: notes on implementation}), then counts down an auxiliary information collection window $\Delta$. The accused parties who fail to provide such information to refute the evidence in time are treated as culpable. We discuss setting $\Delta$ in \Cref{sec: discussion}. 
The accused server(s) open the corresponding commitments and reveal inputs to $f(\cdot)$ (line 9). \textbf{CC} verifies the correct opening of commitments and compares the output of $f(\cdot)$ with the evidence (line 10). 
Note that we do not distinguish between collusion that recovers the desired entry $\D_a$ and random entry $\D_{a_r}$. 

\paragraph{Intuitions behind incentive design} 
The first correct collusion reporting server for a PIR run is selected as the winner. 
The payment rules are as follows: (1) fine servers submitting false reports \amt{f}; (2) distribute rewards \amt{r} to the winner; (3) penalize each server being successfully accused amount \amt{p}; (4) charge service fee \amt{s} for each query and pay a queried server that is not successful accused amount \amt{s}. (1) The framing fines are to discourage servers from false accusations. (2) Rewarding the first reporter is to encourage the servers participating in collusion to reveal its existence in infinitely repeated runs of the PIR service. (3) Fining the accused colluding parties is to discourage at least one party in the colluding group without a network advantage from colluding. (4) Collecting service fees from clients is first to stop clients from exploiting the mechanism and causing undue punishment on servers, and second to determine the length of the privacy protection period, which we discuss in more detail 
in \Cref{sec: exiting}. 

To realize the incentive scheme, in the beginning, servers make deposits greater than \amt{p} and \amt{f} to \textbf{CC}. Deposits function as sources of potential fines and penalties, and will be transferred back when servers exit the service. Users also make deposits \amt{s} before making each query.

\paragraph{Organization of contents}
We next dive into the key analysis for designing the mechanisms that achieve the non-collusion outcome while accounting for manipulations from servers and users in \Cref{sec:realtime}. We add malicious servers \tian{and clients} in \Cref{sec: malicious} and update the collusion evidence verification routines. \Cref{sec: extended} discusses servers' exiting strategies, and how strong coalitions form. \Cref{sec: game demo} summarizes the mechanism and presents an implementation. 

%% file: sections/5_collusion-k-servers.tex
\section{Collusion deterrence}\label{sec:realtime}
As mentioned in \Cref{sec: pir intro}, we base the analysis on the premise that after successful collusion, at least some colluders learn some nontrivial function of the user secrets. 
We now dissect the intrinsic incentive structure of the solution. 
We let all servers be rational and include malicious servers in \Cref{sec: malicious}. 

\subsection{Single-shot game}\label{sec:beta equilibrium}
During a single-shot collusion, colluding servers verify the information input by others, e.g., by including circuits that examine users' signatures in de-commit information. 

\paragraph{Order of events} 
Consider the following single-shot sequential game $\mathsf{G}$: 
\begin{enumerate}
 \item[(1)] In round $1$, servers decide whether to collude, $C$ or $\bar{C}$.
 \item[(2)] In round $2$, colluding servers decide whether to input the correct information, amicable ($A$) or deceitful ($D$). 
 \item[(3)] In round $3$, each server decides whether to report collusion, $R$ or $\bar{R}$. 
\end{enumerate}

\paragraph{Solution concept}
The game we have defined is a sequential game of complete information, i.e., players know others' past actions. Nash equilibria, the standard notion of equilibrium in game theory, are known to predict ``unreasonable'' behaviors in sequential games. In a Nash equilibrium of a sequential game, a player can commit to empty/non-credible threats.\footnote{Consider the Ultimatum game with two players. One player proposes a way to divide a sum of money between them. If the responding player accepts the proposal, then the money is divided accordingly; otherwise, both receive nothing. The responder can threaten to accept only fair offers, but this threat is not credible: given any non-zero amount, the only rational action for the responder is to accept.} 
We then consider a refinement of Nash equilibrium for sequential games, namely \textit{Subgame Perfect Equilibrium}~\cite{selten1988reexamination} (SPE). 
In SPE, a player chooses the optimal action at any point, i.e., the action that maximizes expected utility from that point on, no matter what has happened in the past.

\paragraph{Earning privacy worth}
Let $V$ be the worth of the user secret. 
If the game stops in round one, each queried server only receives service fee \amt{s}. Otherwise, if they all play amicable ($A$) in round two, they all learn the secret and receive $V$. If at least two parties play $D$, then they do not learn the secret. If there is exactly one deceitful party, then others do not learn the secret and the deceitful server learns the secret in its best case, e.g., when colluders exchange queries or responses, but not otherwise.

\paragraph{SPE} 
In solving the collusion game $\mathsf{G}$, we let $\amtt{r}=0$. 
The result for the single-shot $k$-party collusion game is as follows. 
\begin{restatable}[]{theorem}{kacl}
 In the $k$-party collusion game $\mathsf{G}$ in $(\ell, k, 1)$-PIR, the strategy profiles of all $\ell$ servers playing $\bar{C} D$ or $C D$ in rounds one and two are the only two SPEs when $\amtt{p}>0$ and $\amtt{s} + \frac{k-1}{k} \amtt{p} > V$.\label{thm: k agent l server case}
\end{restatable}

\begin{figure*}[ht]
    \centering
    \includegraphics[width=0.56\textwidth]{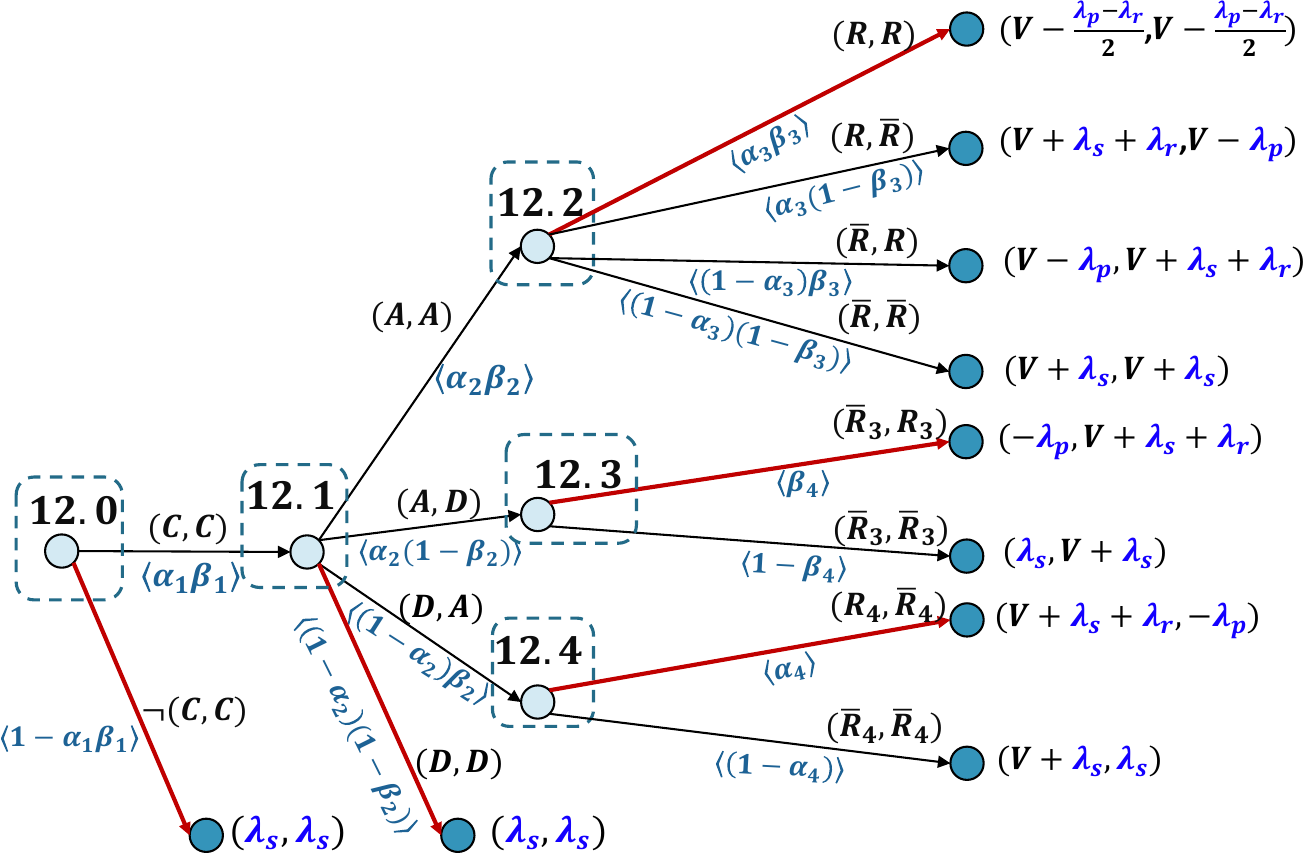}
    \caption{Extensive form of $\mathsf{G}$ with $k=2$. Denote the queried servers as $S_1, S_2$. Each node enveloped with dashed lines is a decision node, with the prefix of its name indicating the owner (e.g., ``12'' means that $S_1$ and $S_2$ make decisions at this node). 
    Actions of servers are specified on the one-directional arrows and the probability for each move is under the arrows in brackets, with $\alpha$s for $S_1$, and $\beta$s for $S_2$. We label the round three actions $R$ and $\bar{R}$ at decision nodes 12.3 and 12.4 respectively as $R_3, \bar{R}_3, R_4, \bar{R}_4$. 
    The payoff vector is in the form $(x, y)$ with $x$ for $S_1$ and $y$ for $S_2$. When both servers report, we apply an equal probability of submitting first. The thick red edges indicate the SPE. 
    }
    \label{fig:demo2servers}
\end{figure*}
In \Cref{thm: k agent l server case}, we only describe the strategies in the first two rounds that are pertinent to our desired non-collusion outcome, where servers do not collude or input gibberish into collusion protocols. 

\begin{proof}
 First, consider $k=2$. The extensive form game tree for 2-server collusion is shown in \Cref{fig:demo2servers}. We depict the good case for the only deceitful player where it learns the secret in \Cref{fig:demo2servers} and consider \emph{both good and bad cases} where it does not learn the secret, in analysis. The bad case can occur if there is a conditional abort upon detection of incorrect inputs. 
 
 At decision node 12.3 and 12.4, it is straightforward to determine $\alpha_4=1$ and $\beta_4=1$ since deviating does not generate more returns when $\amtt{r} = 0$. This means that the deceitful server reports an amicable colluder. 
 At decision node 12.2, to make $R$ rational, we need $R$ to generate higher expected returns:
 \begin{align*}
 R: &\beta_3 (V - \frac{\amtt{p} - \amtt{r}}{2}) + (1- \beta_3) (V + \amtt{s} + \amtt{r}) \\ 
 &= V + \amtt{s} - \beta_2 (\frac{\amtt{p}}{2} + \amtt{s}) \\
 \bar{R}: &\beta_3 (V - \amtt{p}) + (1- \beta_3) (V + \amtt{s}) \\
 & = V + \amtt{s} - \beta_2 (\amtt{p} + \amtt{s})
 \end{align*}
 Playing $R$ produces higher expected returns since $\amtt{p} > 0$. Then $\alpha_3 = \beta_3 = 1$. 
 Working backward, at decision node 12.1, we consider the following:
 \begin{align*}
 \text{Good case: } 
 A: &\beta_2 (V - \frac{\amtt{p} - \amtt{r}}{2}) + (1 - \beta_2) (- \amtt{p}) \\
 & = - (1 - \frac{\beta_2}{2}) \amtt{p} + \beta_2 V \\
 D: &\beta_2 (V + \amtt{s} + \amtt{r} ) + (1 - \beta_2) \amtt{s} 
 = \amtt{s} + \beta_2 V \\
 \text{Bad case: } 
 A: &\beta_2 (V - \frac{\amtt{p} - \amtt{r}}{2}) + (1 - \beta_2) \amtt{s} \\
 &= \beta_2 (V - \frac{\amtt{p}}{2} - \amtt{s}) + \amtt{s} \\
 D: &\beta_2 \amtt{s} + (1 - \beta_2) \amtt{s} = \amtt{s} 
 \end{align*}
 Playing $D$ generates higher returns in good case and at least the same returns in bad case since $\amtt{s} + \frac{\amtt{p}}{2} > V$. Then $\alpha_2 = 0$ and similarly, $\beta_2 = 0$. 
 At decision node 12.0, playing $\bar{C}$ gives the same returns as playing $C$, i.e., service fee \amt{s}. Then $\alpha_1 = \beta_1 = 0$ or $\alpha_1 = \beta_1 = 1$. Hence, the SPEs for $k=2$ are $\bar{C}DR \bar{R}_3 R_4$ and $C D R \bar{R}_3 R_4$. 
 
 \begin{figure*}
 \centering
 \includegraphics[width=0.72\textwidth]{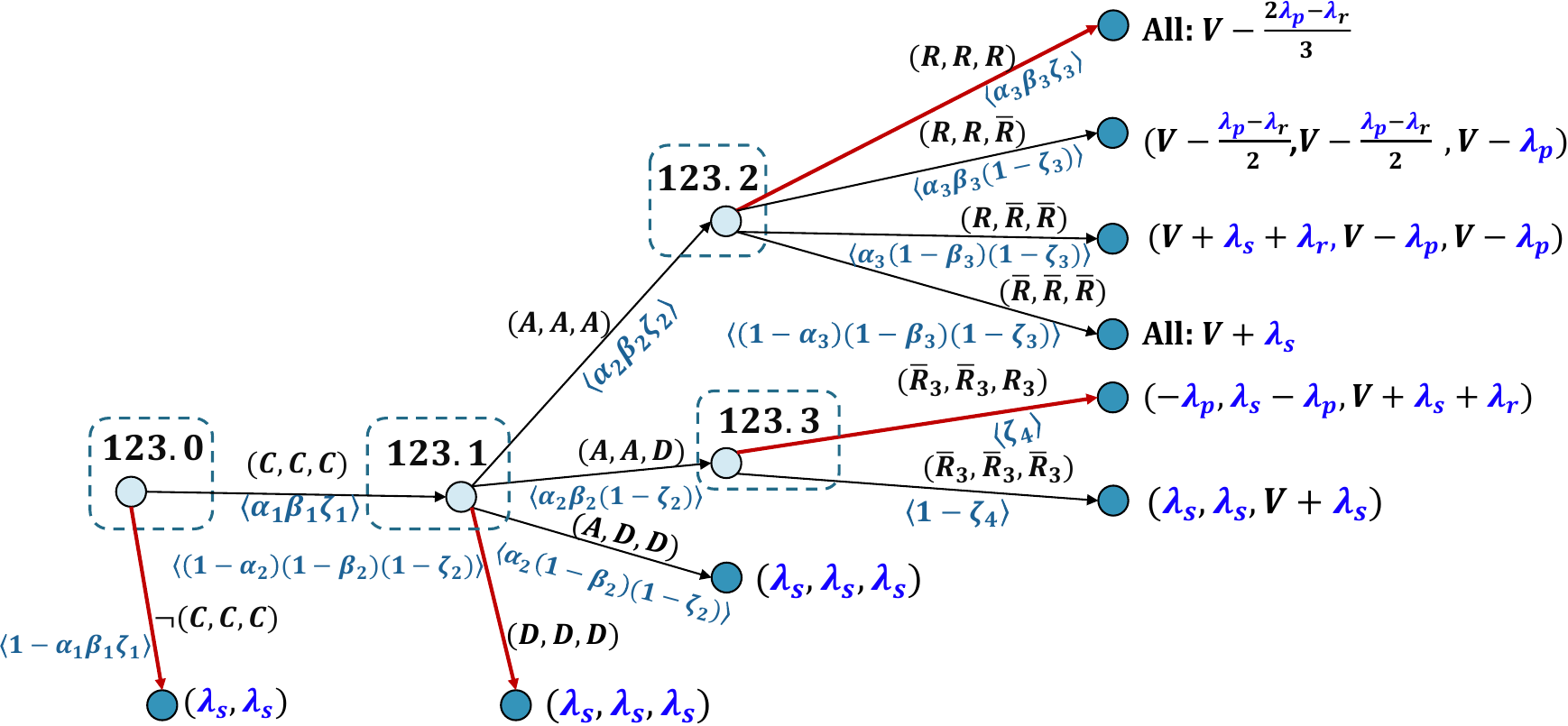}
 \caption{Extensive form of $\mathsf{G}$ with $k=3$. We label the round three actions $R$ and $\bar{R}$ at decision nodes 123.3 as $R_3, \bar{R}_3$. The thick red edges indicate the SPE. 
 }
 \label{fig:demo3servers}
 \end{figure*}
 Next, consider $k>2$. The example extensive form game tree for $k=3$ can be found in \Cref{fig:demo3servers}. 
 In round three, it is straightforward to see that when there is more than one deceitful player, all parties can only play $\bar{R}$. When there is exactly one deceitful player, e.g., decision node 123.3 in \Cref{fig:demo3servers}, this player plays $R$ and others play $\bar{R}$. When all players are amicable, e.g., node 123.2, the expected returns from playing $R$ are greater than playing $\bar{R}$ when $\amtt{p} > 0$ because $V - \frac{k-1}{k}\amtt{p}>V-\amtt{p}$, similar to the $k=2$ case. 
 Next, in round two, let $x_1$ be the probability of exactly one other party playing $D$ ($x_1\in [0,1]$), and $x_2$ be the probability of at least two other parties playing $D$ ($x_2\in [0,1-x_1]$). We can compute the returns from playing $A$ and $D$ as follows:
 \begin{align*}
 \text{Good case: } 
 A: &x_1 (-\amtt{p}) + x_2 \amtt{s} + (1-x_1-x_2) \cdot \\ &(V - \frac{(k-1)\amtt{p} - \amtt{r}}{k}) \\
 D: &x_1 \amtt{s} + x_2 \amtt{s} + (1-x_1-x_2) (V + \amtt{s} + \amtt{r} ) \\
 \text{Bad case: } 
 A: &x_1 \amtt{s} + x_2 \amtt{s} + (1-x_1-x_2) \cdot\\ & (V - \frac{(k-1)\amtt{p} - \amtt{r}}{k}) \\
 D: &x_1 \amtt{s} + x_2 \amtt{s} + (1-x_1-x_2) \amtt{s} 
 \end{align*}
 It is straightforward to see that playing $D$ produces higher expected returns. 
 Then in round one, playing $\bar{C}$ produces the same returns as playing $C$. Therefore, playing $\bar{C} D$ or $C D$ in rounds one and two are the SPEs. 
\end{proof}

\paragraph{Counter false accusations} 
(1) For random accusations, for now, we focus on $f(\cdot)$ that reveals explicit bits about user secrets and consider a more general case in \Cref{sec: discussion}. A server can then make successful random accusations with probability $\frac{1}{2}$. (2) An informed queried server knows some non-trivial function about the user secret. It can falsely accuse others successfully with probability $\frac{\omega+2}{2(\omega + 1)}$ ($=\frac{1}{\omega+1}\cdot 1 + \frac{\omega}{\omega+1}\cdot \frac{1}{2}$). 
Combining the two cases, we require $(1 - \max{\{ \frac{1}{2}, \frac{\omega+2}{2(\omega + 1)}} \} )\amtt{f}>\max{\{ \frac{1}{2}, \frac{\omega+2}{2(\omega + 1)}} \} \amtt{r}$ or simply $\amtt{f} > 0$ so that servers do not falsely accuse others. 
(3) A user can also collude with a queried server to make fake accusations. By having $(k-1) \amtt{s} > \amtt{r} = 0$ or simply $\amtt{s}>0$, the user is discouraged from falsely accusing servers. 
Note that a protocol that defends against framing is also called \textit{non-imputable} or achieving \textit{non-imputability}~\cite{goyal2021traceable}.

\paragraph{Comments on the simultaneous moves}
In round two, the parties select action $A$ or $D$ simultaneously. Alternatively, if simultaneity cannot be implemented, one party has to move first. Then no matter what action this initiator player takes, the follower's dominant strategy is $D$. The SPEs stay the same. We give more advantages to colluders by allowing simultaneous moves.

\paragraph{Blind collusion}
When parties play $D$, without embedded verification gadgets, it is computationally hard to fabricate seemingly valid inputs for many PIR protocols, e.g., the Boyle-Gilboa-Ishai and Hafiz-Henry protocols we consider, while being (non-negligibly) possible for others. We focus on collusion with verification and the computationally hard case without verification in the main body. Appendix \ref{sec:deceitful} addresses the latter blind collusion without verifying inputs. 
There, we solve for the conditions on parameters to achieve the non-colluding sequential equilibrium~\cite{kreps1982sequential} in \Cref{prop: 2 agent l server case}. 

\subsection{Non-collusion outcome in repeated games}\label{sec:repeated}
Consider the following repeated collusion game $\mathsf{G_R}$ for a player $S_i$. In each run $1,\ldots, T$:
\begin{enumerate}
 \item[(0)] In round zero, Nature decides whether $S_i$ is queried. If not, $S_i$ terminates this run and enters round zero of the next run. Otherwise, $S_i$ enters game $\mathsf{G}$. 
\end{enumerate}
When $T$ is finite and known, we can solve for the equilibrium using backward induction as before. In the last run of the game where $S_i$ is queried, regardless of history, the SPEs are the same as before under the same set of conditions, i.e., $S_i$ reports after successful collusion in round three, plays $D$ in round two, and plays $C$ or $\bar{C}$ in round one. With this knowledge, players do not deviate from the prescribed SPEs in the single-shot game in the second-to-last run, and so on, until the first run. We state the following corollary and omit the proof since it follows from \Cref{thm: k agent l server case}.
\begin{corollary}
    The $\ell$-party repeated collusion game $\mathsf{G_R}$ in $(\ell, k, 1)$-PIR with a known finite number of runs has the same SPEs under the same conditions as in \Cref{thm: k agent l server case}.\label{cor: finite repeatition}
\end{corollary}

When $T$ is infinite or finite but unknown, backward induction no longer applies and collusion can become an equilibrium~\cite{mailath2006repeated}. For simplicity, we address both cases as infinitely repeated games. The Folk Theorem~\cite{friedman1971non} states that if players are patient enough, then repeated interaction can result in virtually any average payoff in an SPE. If the players are memoryless (i.e., always play like they are playing for the first time), we can still achieve the non-colluding equilibrium. We show the same result \textit{without} making such additional assumptions. Specifically, we consider player strategies of the form ``If another player has reported collusion, never collude with this player again''. Because strategies of this form yield undesirable equilibria in many repeated versions of well-studied games, e.g., repeated prisoner's dilemma. 

\paragraph{SPE} 
In solving game $\mathsf{G_R}$ with unknown or infinite $T$, we let the reward $\amtt{r} > 0$. 
Same as before and after simplification, we require $ \amtt{f} > \frac{\omega+2}{\omega} \amtt{r}$ 
and $\amtt{s} > \frac{1}{k-1} \amtt{r} $ so that servers and users do not falsely accuse others. 
We now formally state the main result for the $\ell$-party repeated collusion game. We apply discount rate $\delta\in [0,1)$ on future returns~\cite[Chap.~7.2]{myerson1997game} because rational agents prefer acquiring the same returns earlier than later. Higher $\delta$ indicates higher patience. 
\begin{restatable}[]{theorem}{kaclr}
 In the $\ell$-party infinitely repeated collusion game $\mathsf{G_R}$ in $(\ell, k, 1)$-PIR, the strategy profile of all $\ell$ servers playing $\bar{C} D$ or $CD $ in rounds one and two of each run are the only SPEs when $\frac{\delta}{1-\delta} (1 - q ) V <\amtt{r}\leq \amtt{p} $, and $\amtt{s} + \amtt{p} > \frac{1}{k} (\amtt{r} + \amtt{p}) + V$ where $q=\frac{(\ell-k)\cdots (\ell-2k+2)}{(\ell-1)\cdots (\ell-k+1)}$.
 \label{thm: k agent l server repeated}
\end{restatable}
We only describe the strategies in the first two rounds relevant to the non-collusion outcome. 
With the same $k$, when we increase $\ell$, we increase $q$, which expands the feasibility region for \amt{r}. 

\begin{proof}
 When playing $\mathsf{G_R}$, suppose for contradiction servers play $CA\bar{R}$ in equilibrium, i.e., collude without reporting, and they always refuse to collude with those who have reported their collusion in the past. 
 Consider a server $S_1$ that deviates to play $R$ in round three of the collusion game $\mathsf{G}$ in a run. 
 Let $\mathsf{p_{i}}$ be the probability that server $S_1$ can collude with the other $(k-1)$ queried servers in the $i$-th run where $S_1$ is queried ($i\geq 1$), meaning that none of them have reported each other before. 

 First, consider $k=2$. Two queried servers collude only if they have never met before. In the $i$-th run, suppose $S_1$ and $S_2$ are queried. The probability that in one of the previous $(i-1)$ instance, $P_2$ is not queried is $\frac{\ell-2}{\ell-1}$. Then for $k=2$, $\mathsf{p_{i}} = (\frac{\ell-2}{\ell -1})^{i-1}$ since the runs are independent when each client samples servers at random by themselves. 
 
 Next, consider $k>2$. Unlike in the $k=2$ case, the $k-1$ queried servers collude only if they have never colluded with $S_1$ before, even though they may have already met in some runs. This is because collusion may have not happened due to some other queried servers refusing to collude with $S_1$. We can obtain a lower bound on $\mathsf{p_{i}}$ by considering collusion only if the $k$ parties have never met $S_1$ before. 
 This gives $\mathsf{p_{i}} \geq \big [\frac{\binom{\ell-k}{k-1}}{\binom{\ell-1}{k-1}} \big ]^{i-1} = \big [\frac{(\ell-k)\cdots (\ell-2k+2)}{(\ell-1)\cdots (\ell-k+1)} \big ]^{i-1} $. 
 Let $q = \frac{(\ell-k)\cdots (\ell-2k+2)}{(\ell-1)\cdots (\ell-k+1)}$. 
 Then $\mathsf{p_{i}} \geq q^{i-1} $.

 From playing $CA\bar{R}$, $S_1$ earns service fees and the privacy worth in each run: 
 \[
 \amtt{s} + V + (\amtt{s} + V) \sum_{i=1}^\infty \delta^i 
 \]
 From reporting collusion, $S_1$ expects to earn rewards in the runs where colluding partners are strangers and at least the service fees otherwise:
 \[
 \aarnr + \sum_{i=1}^\infty (\mathsf{p_{i+1}} (\aarnr) + (1 - \mathsf{p_{i+1}}) \amtt{s}) \delta^i 
 \]
 Subtracting the first quantity from the second gives 
 \begin{multline*}
 \amtt{r} + (V +\amtt{r}) \sum_{i=1}^\infty \mathsf{p_{i+1}} \delta^i - \vcgmath \frac{\delta}{1-\delta} \geq \\ \amtt{r} + (V +\amtt{r}) \sum_{i=1}^\infty ( q \delta)^i - \vcgmath \frac{\delta}{1-\delta}
 \end{multline*}
 Then, as long as $\frac{1}{1-q\delta} \amtt{r} > (\frac{\delta}{1-\delta} - \frac{q\delta}{1-q\delta} )V$, playing $\bar{R}$ cannot be equilibrium strategy since playing $R$ is a profitable deviation. This is satisfied when $\amtt{r} > \frac{\delta}{1-\delta} (1 - q ) V $. 
 This means that one reports each successful collusion. Then following from \Cref{thm: k agent l server case}, in round two of each collusion game, playing $D$ generates higher returns, and in round one, playing $\bar{C}$ or $C$ generates the same returns. Playing $\bar{C}D$ or $CD$ in the first two rounds of each run are the only SPEs. 
\end{proof}

\paragraph{What is special about infinite repetition}
The theory of infinitely repeated games often faces a multiplicity of equilibria~\cite{dal2011evolution} dependent on players' patience, i.e., how much they discount future returns. 
For example, in a single or finitely repeated two-person prisoners' dilemma, both players playing ``defect'' is the Nash equilibrium; in an infinitely repeated prisoners' dilemma, ``cooperate'' becomes the equilibrium strategy when they are sufficiently patient and discount future returns by a sufficiently small amount. This is because there is no end game, and a player can always penalize ``defect'' actions. 
Except for penalties, in broader social contexts, equilibrium different from those appearing in finite games can also arise due to reputation concerns in long-run relationships and shared expectations of social norms. 
Examples include neighbors in Shasta County, CA, resolving disputes caused by each other's cattle via informal social norms instead of legal rules without any hierarchical coordinator~\cite{ellickson1991order}, and anonymous sellers of a certain size in Deep Web exhibiting honest patterns to amass good reputations without government regulations~\cite{hardy2016reputation}. 

Our goal is then to preserve the equilibrium in finite games in infinite runs, even when players are highly patient. The first step is to obstruct the formation of informal institutions~\cite{helmke2004informal} or long-term relationships among the players by having many servers and querying only some of them in each run. Combining this with positive rewards, we increase players' returns from following the prescribed equilibrium strategy in the current run and decrease the probability of them playing together again. In this way, even a patient player expects higher utility from observing the desired non-colluding strategy. 

\subsection{Existence of solution}
Now one natural question is whether practical solutions are guaranteed to exist. To ensure affordable service fees \amt{s}, we define practical service fees mathematically as $\amtt{s} \leq \frac{V}{k} \xi$, where $\xi\in (0,1)$ is a practicality parameter characterizing the affordability of \amt{s}. We state the following proposition. 

\begin{restatable}[Existence of solution]{proposition}{existence} For $\ell \gg k$, there always exist parameter assignments for inequalities \circno{1}-\circno{5}: \circno{1} $\amtt{f} > \frac{\omega+2}{\omega} \amtt{r}$; 
 \circno{2} $(k-1) \amtt{s} > \amtt{r} $; 
 \circno{3} $\frac{\delta}{1-\delta} (1 - \frac{(\ell-k)\cdots (\ell-2k+2)}{(\ell-1)\cdots (\ell-k+1)} ) V <\amtt{r}\leq \amtt{p} $; 
 \circno{4} $\amtt{s} + \amtt{p} > \frac{1}{k} (\amtt{r} + \amtt{p}) + V$; 
 \circno{5} $\amtt{s} \leq \frac{V}{k} \xi$, where $\delta\in [0,1), \xi \in (0,1)$.
 \label{thm:exist}
\end{restatable} 
\begin{proof}
 To satisfy inequality \circno{2} and \circno{5}, we need $(k-1)\frac{V}{k}\xi > \amtt{r} $, which gives $\xi \geq \frac{k \amtt{r}}{ (k-1) V}$. We can have a feasible $\xi<1$ by mandating $k \amtt{r} < (k-1) V$. Considering \circno{3}, we have feasible assignments for the reward amount if $\frac{\delta}{1-\delta} (1 - (\frac{\ell - 2k + 2}{\ell - k +1})^{k-1} ) < \frac{k-1}{k} $. Then we can pick \amt{f} and \amt{p} according to inequality \circno{1}, \circno{3} and \circno{4}, \amt{s} according to inequality \circno{2}, \circno{4} and \circno{5}. 
\end{proof}

\begin{figure}
 \centering
 \includegraphics[width=0.85\linewidth]{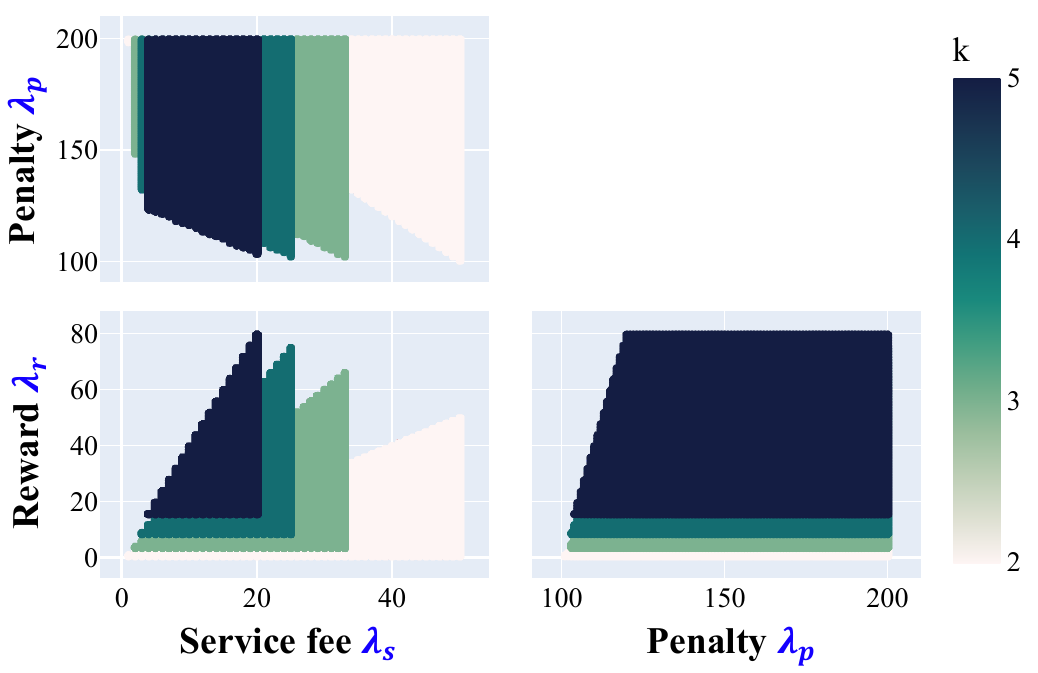}
 \caption{Example parameter feasibility regions. We provide a 3D interactive plot here~\cite{figsrc}.}\label{fig:plot_sai}
\end{figure}

An example feasible parameter region of \amt{s} and \amt{p} in setting $\ell=10000, V=100, k=\{2,3,4,5\},w =1, \amtt{p}\leq 2V, \delta=0.99$ is shown in \Cref{fig:plot_sai}. Higher service fee allows for a lower penalty (\circno{4}) and higher reporting reward (\circno{2}). Lower $\delta$ will expand the feasibility regions (\circno{3}) because players are less patient and prefer receiving the same payment earlier. 
One sample solution here is $k=2,\amtt{s}=1, \amtt{f}=\amtt{p}=200, \amtt{r}=0.99$. 

%% file: sections/7_privacy.tex
\section{Malicious client or servers}\label{sec: malicious}
We now consider \tian{the client} or $\theta\in [0,1)$ of the $\ell$ servers to be malicious. 
Malicious servers behave arbitrarily and ignore incentives. They bring about two changes to the mechanism. First, the incentive mechanism achieves the non-collusion outcome only probabilistically. Second, the collusion evidence verification routines described in \Cref{fig:solution} can be exploited by malicious servers to function as an expensive oracle for revealing user secrets. 

\tian{A malicious client may collude with a server to frame other servers, even if the framing results in \textit{losses}. We have discussed false accusations from rational users and clients in \Cref{sec:beta equilibrium} and touch on framing from malicious clients in this section.}

\subsection{Statistical non-collusion outcome} 
Consider a statistical security parameter $\eta$, e.g., $\eta=40$. 
To achieve the non-collusion outcome with probability at least $1-2^{-\eta}$, $\theta$ needs to satisfy 
 \[
 \frac{\binom{(1-\theta)\ell}{0} \binom{\theta \ell}{k}}{\binom{\ell}{k}} + \frac{\binom{(1-\theta)\ell}{1} \binom{\theta \ell}{k-1}}{\binom{\ell}{k}} \leq 2^{-\eta} \tag{$S^\star$}
 \] 
so that at least two of the queried servers are rational. 
Following from \Cref{thm: k agent l server case} and \ref{thm: k agent l server repeated}, we have the following corollaries.
\begin{corollary}
 In the $k$-party collusion game $\mathsf{G}$ in $(\ell, k, 1)$-PIR with $\theta \ell$ malicious servers, with probability $1-2^{-\eta}$, the strategy profile of all $(1-\theta)\ell$ rational servers playing $\bar{C} D$ or $CD$ in rounds one and two are the only SPEs when $\amtt{p}>0$ and $\amtt{s} + \frac{1}{2} \amtt{p} > V$ where $\theta$ satisfies $S^\star$.\label{cor: k agent l server malicious case}
\end{corollary}

\begin{corollary}
 In the $\ell$-party repeated collusion game $\mathsf{G_R}$ with an infinite or unknown number of runs in $(\ell, k, 1)$-PIR with $\theta \ell$ malicious servers, with probability $1-2^{-\eta}$, the strategy profile of all $(1-\theta)\ell$ rational servers playing $\bar{C} D$ or $CD$ in rounds one and two of each run are the only SPEs when $\frac{\delta}{1-\delta} (1 - q ) V <\amtt{r}\leq \amtt{p} $, and $\amtt{s} + \amtt{p} > \frac{1}{2} (\amtt{r} + \amtt{p}) + V$ where $q=\frac{(\ell-k)\cdots (\ell-2k+2)}{(\ell-1)\cdots (\ell-k+1)}$ and $\theta$ satisfies $S^\star$.
 \label{cor: k agent l server repeated malicious}
\end{corollary}

\subsection{Privacy-preserving verification}
The second change is that now we need to protect user secrets during collusion resolution. This is not a concern when servers are rational since they do not collude in equilibrium. But malicious parties do not respond to incentives, they can falsely report collusion and learn the secret by paying the penalty \amt{p}. 
Therefore, in collusion verification, instead of having involved parties reveal commitments and perform verification on plaintexts, we adopt zero-knowledge accusation verification.

Upon receiving an accusation with a nontrivial $f(\cdot)$, \textbf{CC} counts down a proof collection window. 
The accused parties collectively provide a zero-knowledge succinct non-interactive argument of knowledge (zkSNARK) $\pi$ justifying the evidence being false (line 2 in \Cref{fig:solution2}). \textbf{CC} confirms the accusation and executes payments if the accused servers fail to provide proof in time $\Delta^*$ or if the proof is rejected (line 3). We discuss setting $\Delta^*$ in \Cref{sec: discussion}. If the accuser somehow composes a zkSNARK $\pi^*$ for the knowledge of inputs and outputs of the function $f(\cdot)$ during the executions of the collusion protocol, $\pi^*$ can be submitted as evidence and the contract checks $\pi^*$ directly.

\begin{figure}[ht]
 \centering
 \begin{mdframed}
 \centering
 \textbf{Zero-knowledge Collusion Resolution ($S_{id_i}$ accusing)}
 \begin{flushleft}{\small
 \begin{enumerate}[wide, labelindent=0pt]
 \item Server $S_{id_i}$ submits a collusion report to \textbf{CC}, indicating the evidence, i.e., circuits for computing $f(\cdot)$, $S_{id_i}$'s inputs to compute $f(\cdot)$, and the commitment to the function outputs. \textbf{CC} rejects the report if $f(\cdot)$ is trivial. 
 \item Accused server(s) provide $\pi$ proving that the committed value in the evidence does not match the function output of $f(\cdot)$. 
 \item \textbf{CC} confirms the accusation if (1) the accused server(s) fails to submit the proof in time $\Delta^*$, or (2) $\pi$ is rejected. 
 \end{enumerate}}
 \end{flushleft}
 \end{mdframed}
 \caption{Overview of zero-knowledge collusion resolution.}
 \label{fig:solution2}
\end{figure}

\paragraph{Constructing zkSNARKs} 
When the evidence in the accusation is a query or response that has been committed to on the board \textbf{BB}, the server being accused can submit \textit{proof of inequality}. If the accuser does not specify the accused, all other servers queried in this instance need to provide inequality proofs. More concretely, suppose we work with Pedersen commitment with common generators $g,h$ and modulus $N$. It is perfectly hiding and satisfies our needs. Let $c=g^m h^r$ be a commitment sent by a party $S_x$ on \textbf{BB} where $m$ is the committed message and $r$ is the randomness. A party accuses with evidence $(g^s, S_x)$, i.e., ``the exponent of $g^s$ is the committed value in $c$'', instead of revealing $s$ directly. Then $S_x$ needs to prove the statement that $c'=c/g^s$ is not a commitment of zero, i.e., proof of knowledge $PK \{(a,r): c'= g^a h ^r \wedge a \neq 0 \wedge r \neq 0\}$ where $a=m-s$. This can be efficiently proven with range proofs~\cite{bunz2018bulletproofs}.

Next, consider the scenario where the evidence is a general function circuit along with its output. There are two scenarios. First, if $k=2$, the accused server generates the proof on its own. Suppose we still work with Pedersen commitment with parameters $g,h, N$. Now suppose a party accuses with general evidence stating that ``the exponent of $g^s$ is the output of $f(\cdot)$'' along with its collusion input $m_1$, which is committed on \textbf{BB} in a commitment $c_1$. Let the accused server hold the corresponding commitment $c_2=g^{m_2} h^r$. She proves $PK \{(m_2,r): c_2 = g^{m_2} h ^r \wedge g^a = g^{f(m_1, m_2)} \wedge g^a \neq g^s\}$. 
For simplicity, we let $f(\cdot)$ be expressed as an arithmetic circuit since the accused party can arithmetize it if $f(\cdot)$ is a Boolean circuit. We can then employ applicable schemes including Groth16~\cite{groth2016size}, Plonk~\cite{gabizon2019plonk}, or LegoSNARK~\cite{campanelli2019legosnark} with a polynomial commitment. 
Second, when $k>2$, the accused servers can utilize the collaborative zkSNARKs primitive~\cite{ozdemir2022experimenting} to generate the proof while preserving privacy for their inputs. The knowledge to prove is similar to before, except that the inputs to $f(\cdot)$ are held by multiple parties.

\tian{
\subsection{Malicious client}
Given a malicious client who knows the queried index and is indifferent to incentives, it is crucial to ensure that the query and answer strings are unknown to the client. 
A straightforward approach is to run the query generation and reconstruction algorithms with MPC as in traceable secret sharing\cite{goyal2021traceable}, where the randomness in the query generation algorithm is generated by the client and all queried servers. 
In this way, the client can successfully retrieve the outputs without learning the query or answer strings.
}

\section{Exiting strategies and coalition formation}\label{sec: extended}
In this section, we analyze how servers' exiting strategies and strong coalition formation affect the proposed mechanism and its parameterization. 

\subsection{Exiting strategy}\label{sec: exiting}
We first study the two exiting strategies, carrying out massive collusion before exiting and performing collusion or data selling after leaving the system.

\paragraph{A server $S_e$ is leaving the system}
The maximum fines we can take from $S_e$ is the deposited penalty \amt{p}. 
We propose a Self-Insurance \cite{ehrlich1972market} design tailored to our needs and detain service fees as the self-insurance pool for a certain privacy protection period of $T$ time units. Suppose during the period, $\Omega$ users submit $k \Omega$ total queries. The accumulated service fees are $k \Omega \amtt{s}$. The expected number of queries that $S_e$ receives is $\frac{k \Omega }{\ell}$. Then $S_e$ can recover $\frac{k \Omega}{\ell}$ user secrets in expectation in its best case. 
Consider the worst case (for us) where all collusion is successful. 
We expect $(k-1)/k$ of the collusion to be reported by colluding servers other than $S_e$. The expected total amount of fines we cannot realize from $S_e$ is $( \frac{k \Omega}{l}\frac{k-1}{k} - 1 ) \amtt{p} = \frac{(k-1) \Omega - \ell }{\ell} \amtt{p}$. 
In the meantime, the service fee pool also grows with $\Omega$. There are $k\cdot \frac{k \Omega}{\ell} \amtt{s}$ transaction fees that are available to cover the potential loss from $S_e$'s exiting. We use \[ \sigma = \frac{\frac{(k-1) \Omega -\ell}{\ell}\amtt{p}}{k \frac{k \Omega}{\ell} \amtt{s}}=\frac{(k-1) \Omega -\ell }{k^2 \Omega}\frac{\amtt{p}}{\amtt{s}} \] 
to denote the tension between the size of inexecutable fines and the service fee pool, and $\sigma\leq 1$ is needed to cover fines with detained service fees. $\ell$ being significantly larger than $k$ can allow a small fee amount $\amtt{s}$. 

To be more practical, we can assume an interest rate $r\in [0,1]$ per time unit on these accumulated service fees, which then become $k \Omega \amtt{s} (1+r)^T$. We can set the privacy protection period to be $T$ and discount the penalty across time by rate $r'\in [0,1]$, which means fewer fines as a query becomes ancient for a positive $r'$. We then obtain \[ \sigma' = \frac{(k-1) \Omega - \ell}{k^2 \Omega} \frac{\amtt{p} (1-r')^T}{ \amtt{s} (1+r)^T}\] 
where $\sigma'\leq 1$ is also desired to relieve the tension. As an illustration, let $k=2,\ell =1000,T=10000,\Omega=5000$. We have $\sigma'=\frac{1}{5}\frac{\amtt{p} (1-r')^{10000}}{ \amtt{s} (1+r)^{10000}}$. For $r'=r=0.0001$, we can have $\sigma'\leq 1$ when $\amtt{s}\geq 0.027 \amtt{p}$. Note that this is not the exact formulation as we are treating the queries as coming at once at the beginning. 
We provide a detailed derivation and consider more servers exiting in \Cref{sec: insurance}.

\paragraph{A server has left the system}
After a server has exited the service, it can sell gathered information to parties inside or outside the system without being penalized. This is outside the scope of this work, but we can enforce the rule that servers only receive their deposits back after the privacy protection period ends for their last received query. 
Additionally, the proof of secure erasure (PoSE) technique~\cite{perito2010secure} that ensures the erasure of memory can potentially counter this threat without keeping servers waiting. We can require a PoSE attestment upon a deposit withdrawal request.

\subsection{Strong coalitions}\label{sec: coalition}
We now extend the analysis to consider strong coalitions formed by servers. The threat model for group players is similar to the individual server case but has three major changes. (1) Servers inside the coalition act in a highly coordinated way. (2) Information entering into the group is shared among members, and (3) members do not report collusion inside. 
 
\subsubsection{Bound the loss within the current model}
Suppose the pre-established coalition is of size $s$, and that the number of servers to be queried by the user is $k $ ($\leq \ell $). We consider the following two cases. 
First, if $s < k-1$, at least two external rational servers are queried. Collusion with $2$ rational outsiders is discouraged by the current mechanism. 

Second, if $s \geq k-1 $, with some probability $p_{s}$, the clique receives at least $k-1$ of the queries. We can calculate $p_{s} = \frac{\binom{s}{k-1} \binom{\ell - s}{1} + \binom{s }{k}}{\binom{\ell}{k}}$. Then the expected loss from privacy breach when $s \geq k-1$ is $p_s \vcgmath$. 
Note that we do not explicitly include malicious servers in coalitions. But they can be incorporated by conceptually adding $\theta \ell$ more members into the coalition. 

\subsubsection{Update the model}\label{sec:trustovertime}
Cooperative game theory~\cite{branzei2008models} studies the formation of coalitions. Our goal is to cap the coalition size. A cooperative game can be formalized with the set of $\ell$ players $S=\{S_i | i\in [\ell] \}$, each nonempty set of coalitions, and a characteristic / value function $v$ mapping a coalition to its collective payoff, $v:2^S \mapsto \mathbb{R}$. For example, the grand coalition has value $v(S)=\ell \vcgmath + k \amtt{s}$. An allocation $\mathsf{x} \in \mathbb{R}^\ell$ for a coalition specifies how to divide the coalition value among members.

\paragraph{Solution concept}
Suppose the coalition structure (partition of $S$) is non-overlapping. We solve for the allocation in the core of any coalition $C$ of size $s$, denoted as $v(C)$. 
A core allocation is a feasible allocation that no other sub-coalition ($\subset C$) can improve on (i.e., give a strictly higher payoff for all). Here, feasibility simply requires that one does not over-allocate the coalition value. For any $C$ with size $s\geq k$, its value function is $(p_s\cdot s \vcgmath + k/\ell \cdot s \amtt{s} )$. 
Then it is straightforward to see that its core is allocating the coalition value evenly: $\{\forall S_i\in C, \mathsf{x}(S_i) = p_s \vcgmath + k /\ell \cdot \amtt{s} \}$. This leads to the grand coalition being the most desired. 

Alternatively, consider a general value function where the coalition $C$ earns $v(C)=p_s D(s) + k/\ell \cdot s \amtt{s} $ where $D(s)$ is the total value earned by group $C$. We let $D(s)$ be non-increasing in $s$. This is reasonable when the user secrets are rivalrous where a party values secrets more when fewer parties also learn, e.g., upcoming valuation changes of a company in the stock market and hidden treasures. 
In such scenarios, individual profits that can be capitalized decrease as more parties learn this information.

First, consider constant coalition value $D(s) = c\vcgmath$ where $c>0$ is a constant. In this setting, the core is $\{\forall S_j\in C, \mathsf{x}(S_j) =p_s / s \cdot c\vcgmath + k/\ell \cdot \amtt{s} \}$. 
By including one more member to $C$, one existing member's extra gain is $(p_{s+1} / (s+1) - p_s / s) \cdot c\vcgmath$. It is straightforward to see that the quantity is always negative when $k=2$. This indicates that the coalition does not grow in size for $k=2$. By setting $k=2$, we can limit coalition size to $2$. 
For $k\geq 3$, by making the extra returns negative, we can obtain the maximum coalition size $\lfloor \frac{(k-2)(k\ell - k +1)}{(k-1)^2} \rfloor$. 
This means that the coalition is incentivized to include new members until it reaches this maximum size. 

Similarly, consider linear depreciation of coalition value with respect to coalition size, i.e., $D(s) = c\vcgmath/s$. The core is $\{\forall S_j\in C, \mathsf{x}(S_j) =p_s / s^2 \cdot c\vcgmath + k/\ell \cdot \amtt{s} \}$. Adding one more member to $C$ increases $p_s$ while reducing privacy gain. An existing member's extra profit from including a new member is $(p_{s+1} / (s+1)^2 - p_s / s^2) \cdot c\vcgmath $. For $k=2$, this quantity is negative, and the coalition does not have an incentive to include new members. By setting $k=2$, we limit coalition size to $2$. 
For $k>2$, we can bound the size of the coalition by enforcing negative extra profits as before. 
Let $c_1 = -(k-1)(k-2), c_2 = (\ell - 2) k^2 - (3\ell-7)k - 5, c_3 = (k-2)(k\ell - k + 1), \Delta = c_2^2 - 4c_1 c_3$, then the coalition includes new members until its size reaches $\lfloor \frac{-c_2 - \sqrt{\Delta}}{2c_1}\rfloor$. We provide a simple graphical demonstration for locating the upper bound of coalition size here~\cite{coalition}. 
One can continue to consider a higher value depreciation rate, which allows for further limiting coalition sizes for $k>3$. 

%% file: sections/6_implementation.tex
\section{Game design flow and implementation}\label{sec: game demo}
A system designer first evaluates the participating servers and determines a recommended $k$. 
She then decides the privacy value upper bound $V$ to protect. One can compute the parameters for the collusion game according to inequalities in \Cref{thm:exist}. 
It outputs a list of feasible value assignments for parameters. There is no absolute standard to select one assignment over another but rather depends on the priorities.

For illustration purposes, we present abstract essential functions for implementing the game and payment rules in \Cref{alg:coordinator}. In Line 15, a user submits queries to a list of servers. The coordinator initializes parameters for future possible accusation resolutions. In Lines 21-24, an accusation is filed, and we call the accusation validation routine. Line 4 checks the validity of the accusation. 
Line 5 calls the payment execution routine to realize payment rules.

\begin{algorithm}[ht!]
 \small
 \caption{\textbf{Coordinator (Essential Functions)}}\label{alg:coordinator}
 \KwIn{$\amtt{r}, \amtt{p}, \amtt{s}, \amtt{f}$}
 \SetKwProg{Fn}{Function}{\string:}{}
 \SetKwProg{FnD}{On new}{\string:}{}
 \SetKwFunction{AccusationValidation}{$\mathsf{AccusationVal}$}
 \SetKwFunction{AccusationValidationUtil}{$\mathsf{ValidationUtilityFunc}$}
 \SetKwFunction{PaymentExec}{$\mathsf{PaymentExec}$}
 \SetKwFunction{Main}{}
 \Fn(
 ){\AccusationValidation{$witness, id, \mathsf{ed}$}}{
 \For{$pk\in \mathsf{Journal}[id].pkList$}{
 \If{$!\mathsf{alreadyFined}[id][pk]$ and $\mathsf{ed}$ is not recorded in $\mathsf{Journal}[id].filed[pk] $}{
 \eIf{\AccusationValidationUtil{$pk,id, \mathsf{ed}$}}{
    $\triangleright$ Record revealed valid evidence \\
    $\mathsf{Journal}[id].filed.append(witness, \mathsf{ed}) $  \\
    \PaymentExec{$witness,id$} \\
 }{
    Take fines \amt{f} from $\mathsf{Journal}[id].witness$\\
 }
 }
 }
 }
 \Fn(
 ){\PaymentExec{$witness,id$}}{
    Read from $\mathsf{Journal}[id]$ the request $sender,pkList$\\
    \For{$pk\in pkList \ \wedge \ pk!=witness$}{
        Take penalty \amt{p} from $pk$\\
        $\mathsf{alreadyFined}[id][pk]=True$\\
    }
    Distribute reward \amt{r} to $witness$\\
 }
 \FnD{incoming message $\mathsf{m}$}{
 \If{$\mathsf{m}$ is a new query}{
    Lock \amt{s} from $\mathsf{m}.sender$ for each $pk\in \mathsf{m}.pkList$\\
    Issue a unique identifier $id$ for the request\\
    $\mathsf{Journal}[id] \leftarrow (\mathsf{m}.sender, \mathsf{m}.requests, \mathsf{m}.pkList)$\\
    \For{$pk \in \mathsf{m}.pkList$}{
        $\mathsf{alreadyFined}[id][pk]=False$\\
    }
 }
 \If{$\mathsf{m}$ is an accusation}{
    Lock \amt{f} from accuser $\mathsf{m}.sender$\\
    \AccusationValidation{$\mathsf{m}.sender, \mathsf{m}.id, \mathsf{m}.\mathsf{ed}$}
 }
 }
\end{algorithm}
 
\subsection{Implementation}\label{sec: notes on implementation}
\paragraph{Smart contract implementation of \textbf{CC}} We treat Ethereum as a public bulletin board and conveniently implement the coordinator contract as a smart contract on Ethereum. The source code in Solidity is made available~\cite{codesrc}. 
The contract maintains the complete life cycle of PIR service and resolves collusion accusations. For evidence verification, as shown in \Cref{fig:solution}, the accuser submits the evidence, followed by the involved server(s) submitting auxiliary information. If no supplementary data is submitted in time, the accusation is automatically marked as successful. The accused server can be pinpointed through elimination in this case. After collecting information, the verification algorithm computes certain function circuits after verifying function non-triviality and input validity. Any secure commitment and signature schemes suffice, e.g., cryptographic hash functions like SHA-3 as commitment scheme and ECDSA signatures~\cite{johnson2001elliptic}. 

To facilitate evidence verification for a general $f(\cdot)$, we provide a sample reconstruction function as a default $f(\cdot)$ inside the contract. So the accuser can omit the step of providing a function circuit if $f(\cdot)$ is the reconstruction algorithm. Otherwise, we ask the accuser to address the function in a pre-determined way, e.g., $f(bytes \; memory[k])$, when providing the circuit for a simpler calling convention. 
We summarize the gas costs of contract deployment and each function call in \Cref{tab:coordinatorcost} in \Cref{sec: notes on implementation}. 

The gas costs of one-time contract deployment and function calls are summarized in \Cref{tab:coordinatorcost}, which we find to be acceptable, e.g., posting requests cost about $0.74$ dollars~\cite{gassrc}. Here, $\mathsf{CheckCircuits(\cdot)}$ examines the non-triviality of functions mentioned in accusations, and ``+'' represents payments to oracle services (discussed below). $\mathsf{VerifyExchange(\cdot)}$ verifies collusion when the evidence is a de-committed query or response. $\mathsf{VerifyGeneralFunc(\cdot)}$ verifies collusion for a general $f(\cdot)$. $\mathsf{zkVerify(\cdot)}$ performs zero-knowledge verification when malicious servers exist (discussed in \Cref{sec: malicious}).
The costs induced by the base PIR service, i.e., fair exchange of service and fees, are approximately the ``Normal service'' functions. The costs added by the mechanism are transmitting fixed-length commitments of the random companion queries and the functions for ``Collusion resolution''.

\begin{table*}[ht]
 \centering
 \caption{Cost estimates in Gas and dollars~\cite{gassrc}. 
 }
 \label{tab:coordinatorcost}\label{tab:gas}
 \begin{tabular}{lcclcc}
 \toprule
 Normal service & Gas & Dollars & Collusion resolution & Gas & Dollars \\ 
 \midrule
 Contract deployment & $4697299$ & \$8.63 & $\mathsf{Accuse(\cdot)}$ & $223766$ & \$0.41 \\
 $\mathsf{Deposit(\cdot)}$ & $105436$ & \$0.19 & $\mathsf{CheckCircuits(\cdot)}$ & $66991$+ & \$0.12+ \\
 $\mathsf{PostRequests(\cdot)}$ & $405657$ & \$0.74 & $\mathsf{VerifyExchange(\cdot)}$ & $61822$ & \$0.11 \\
 $\mathsf{SubmitResponse(\cdot)}$ & $97400$ & \$0.18 & $\mathsf{VerifyGeneralFunc(\cdot)}$ & $275279$ & \$0.51 \\
 $\mathsf{ClaimServiceFee(\cdot)}$ & $33103$ & \$0.06 & $\mathsf{zkVerify(\cdot)}$ & $2286423$ & \$4.20 \\ 
 \bottomrule
 \end{tabular}
\end{table*}

\paragraph{Cryptographic, communication and computation overhead} 
We preserve the communication complexity of the underlying multi-server PIR because participants only send an additional commitment and signature for each message. 
Computing SHA-3 hash costs 30 base gas and 6 gas every 256 bits~\cite{yellow}, and verifying ECDSA signatures involves recovering the signer with $\mathsf{ecrecover}$, which consumes 3000 gas. 
In terms of computation overhead, SHA-3 is efficient, and an ECDSA signature can be computed in about 0.1ms~\cite{ecdsacmp}. This delay in generating responses is not significant. In two efficient single-server cPIR solutions, XPIR~\cite{aguilar2016xpir} and SealPIR~\cite{angel2018pir}, it takes a server three orders of magnitude more time to compute the response even for small databases ($N=2^{16}$). 

Note that the gas costs related to verifying signatures and messages are not induced by collusion mitigation, but to counter malicious behaviors of sending incorrect messages in general. 
Existing methods for combating the problem of malicious server sending incorrect messages brings even higher inefficiency: 
naively applying GMW compiler~\cite{goldreich2019play} invokes higher communication costs than sending the database; robust multi-server solutions~\cite{goldberg2007improving,devet2012optimally,beimel2007robust} are not generally applicable to any existing schemes, do not support database update, and also have high communication complexity; verifiable PIR~\cite{zhang2014verifiable} relies on cryptographic assumptions and requires a trusted third party to generate verification keys.

\paragraph{Non-triviality of function circuits} 
When parties report with evidence for a general $f(\cdot)$, we examine the non-triviality of the function $f(\cdot)$ in addition to verifying function inputs and outputs. 
We view a function as non-trivial if its output depends on inputs, and taint analysis~\cite{newsome2005dynamic} can be employed for this purpose. 
When we implement the design on Ethereum with smart contracts, the functions do not contain pointers and are deterministic. 

However, accomplishing this task on-chain can be impractically expensive since it essentially requires compiling smart contracts on-chain. Therefore, we treat the function circuit provided by the accuser as non-trivial by default and only verify its non-triviality if the accused indicates its triviality before providing auxiliary information. 
To verify, \textbf{CC} creates an oracle contract and a job verifying the non-triviality of smart contracts on an oracle service, e.g., Chainlink~\cite{breidenbach2021chainlink}. \textbf{CC} creates requests through API calls to the oracle and receives responses. 
Oracle nodes can perform off-chain control and data flow analysis on EVM (Ethereum Virtual Machine) bytecode of smart contracts via EthIR~\cite{albert2018ethir} and Securify~\cite{tsankov2018securify}, among other tools. 
The costs are initially paid by the complainant and charged to the accuser if the function is indeed trivial. 

\paragraph{Zero-knowledge coordinator}
We implement a privacy-preserving coordinator at the same source~\cite{codesrc} and additionally demonstrate a verifier contract for the special information exchange function with Circom compiler~\cite{circom}. Plonk~\cite{gabizon2019plonk} is employed for proof generation and verification. 

\subsection{Discussion}\label{sec: discussion}

\paragraph{Limitation of supported functions}
Overall we prefer $f(\cdot)$ to be light computations, e.g., the entire or the most significant bit (MSB) of the reconstructed data entry. Because it is uneconomical to verify computation-intensive functions when the privacy worth is not high enough. But we also consider this requirement to be undemanding. No matter what the colluding parties have computed during collusion, at least some party, with arbitrary private knowledge, learns some bit(s) or a fraction of a bit about the queried entry from the output. Otherwise, we consider the collusion to be not successful. If a party learns some bits, it can generate a simple explicit function for verification purposes. 

Note that although we focus on $f(\cdot)$ that reveals explicit bits about the index or entry, it is feasible to consider arbitrary information gain, e.g., learning that the index is not $1$. We only need to update inequality \circno{1} in \Cref{thm:exist}, which counters false accusations from informed servers. Suppose servers guess the arbitrary information gain correctly with probability $\leq \tilde{p}$. Inequality \circno{1} is updated to $(1 - \max{\{ \tilde{p}, \frac{\tilde{p} \omega+1}{\omega + 1}} \} )\amtt{f}>\max{\{ \tilde{p}, \frac{\tilde{p} \omega+ 1}{\omega + 1}} \} \} \amtt{r} $, i.e., $\amtt{f} > \frac{\tilde{p} \omega + 1}{(1- \tilde{p}) \omega} \amtt{r}$. 
This means that for larger databases, the small information gain of learning that an index is not a specific one requires higher deposits to discourage false accusations. In such applications, the designer can make a trade-off between the smallest information gain to capture and reasonable \amt{f}.

\paragraph{Set $\Delta$ and $\Delta^*$}
If the auxiliary information and proof collection time, $\Delta$ and $\Delta^*$, are too small, an accuser can initiate denial-of-service (DoS) attacks on the accused servers. When the accused cannot provide auxiliary information or proof in time, the accuser can force a successful false accusation. 
To set $\Delta$, we capture the maximum allowed offline time $\delta_1$ and the downtime $\delta_2$ that can be caused by DoS attacks. We then let $\Delta > \max \{\delta_1, \delta_2\}$. 
To set $\Delta^*$, we let it be larger than $\delta_1$ and $\delta_2$ plus the proof generation time. For proof generation, the least efficient case is multiple accused parties generating proofs with collaborative zkSNARKs. This takes only $\approx 488 \mu s$ per constraint for two parties~\cite{ozdemir2022experimenting}. 

To decide $\delta_1$, we can follow the convention in blockchain systems and let it be a few weeks. For example, hashlocked time contracts in Lightning Network have a maximum delay of $14$ days~\cite{lightningcltv}; 
Ethereum imposes offline penalties to motivate nodes to stay online, e.g., validators lose \textasciitilde40\% balance for inactivity of \textasciitilde36 days\cite{inactivityfines}. 
To decide $\delta_2$, network-layer DDoS attacks empirically last for short periods, e.g., Cloudflare reports 99\% of such attacks last under $3$ hours (and 86\% under $10$ minutes) in the first quarter of 2023~\cite{cloudflarestats}. We can put a conservative estimate of a few weeks for $\delta_2$.

\paragraph{Accommodate small $\ell$ in infinitely repeated PIR service}
To allow $\ell=k$, we can completely replace servers at designated time intervals. Given a finite known service deadline, we achieve the non-collusion equilibrium by \Cref{cor: finite repeatition}.

%% file: sections/9_literature.tex
\section{Related work}\label{sec:literature}
In distributed protocols, participants can collude to compromise various aspects of security, including breaking privacy in MPC~\cite{goldreich2019play}, causing equivocations in state machine replication~\cite{schneider1990implementing}, among many others. 
This work examines privacy-targeted collusion in PIR, and it is relevant to problems with privacy guarantees including generic MPC. 

\paragraph{Mask privacy-targeted collusion} 
Clarke et al.~\cite{clarke2001freenet} circumvent collusion problem in private data retrieval through anonymization. As a result, a privacy breach is not bound to an identified entity. It is debatable whether this suffices as an effective means of protecting privacy. Besides, a user needs to function as a node in the network, making it unsuitable for some applications including PIR.

\paragraph{Mitigate privacy-targeted collusion assuming rationality}
Mangipudi et al.~\cite{mangipudi2021} considers collusion-deterrent information escrow service allowing users to encrypt a message that is only revealed once a pre-specified condition is satisfied. Servers can collude to recover the user message prematurely, but this causes a secret key that can transfer others' deposits to be revealed. There are three major differences from our approach. (1) First, in PIR, servers need to perform computations on queries sent by users, and encrypting messages in special manners is not compatible with PIR constructions that we consider. (2) Second, we allow servers to have arbitrary prior knowledge and capture information gain smaller than learning the entire query or response, i.e., a nontrivial function $f(\cdot)$ on the user secret. Combined with a finite database, random and false accusations become active concerns. (3) In our setting, individual servers earn the entire worth of user secrets instead of splitting the value $V$ among colluders. Besides, we consider repeated runs of the service and malicious parties. 

\paragraph{Deter correctness-targeted collusion} 
Previous works have explored ensuring correct executions by enforcing accountability~\cite{haeberlen2009fault,civit2021polygraph,sheng2021bft}, where at least some misbehaving parties can be located in case of violations. We refer the readers to pertinent surveys and focus on the game-theoretic approaches here. 
Two main game-theoretic collusion mitigation techniques have been proposed, \textit{slashing} non-reporting collusion participants~\cite{dong2017betrayal,yakira2019rational} and having \textit{undercover police} send fake collusion requests to catch colluders~\cite{wang2016information}. 

For the first method, Dong et al.~\cite{dong2017betrayal} focus on ensuring correctness in replication-based cloud computing and propose a solution involving inducing betrayal among colluding parties. 
The four major differences from mitigating \textit{privacy-targeted collusion} are that (1) first, the correctness-targeted collusion affects protocol executions, thus explicitly leaving observable traces and making collusion evidence well-defined. (2) Privacy-targeted collusion can happen (repeatedly) long after the protocol has terminated. (3) False accusations from parties with information advantages are not a concern there. But in privacy-targeted collusion, servers may hold arbitrary private knowledge, which needs to be distinguished from learning user secrets via collusion. (4) Privacy protection during collusion deterrence is naturally not required there. Besides, computing parties there collude via a smart contract, and we allow the collusion process to be unobserved. 

Yakira et al.~\cite{yakira2019rational} presents a slashing mechanism for collusion mitigation in threshold cryptosystems. All participants initially register in an escrow service. One can frame colluding players by submitting proper evidence to the service, which slashes all other agents by burning their deposits and rewarding the reporting agent. There, privacy leakage and active false accusations are not a concern, unlike in PIR where there is a finite number of entries. Besides, the evidence and its verification are not discussed. 

Ciampi et al.~\cite{ciampi2020collusion} present collusion preserving secure (CP-secure) computation protocol, utilizing a collateral and compensation mechanism to disincentivize parties from aborting. Parties initially deposit collaterals and can withdraw only if all messages and executions are correct. Otherwise, the collateral is taken from dishonest players to reward others. In this design, correct messaging and executions are also well-defined and self-contained. 

In the second approach, Wang et al.~\cite{wang2016information} consider collusion deterrence in MPC. One game-theoretic approach proposed there and also in~\cite{wang2014efficient} is to have undercover agents disguised as corrupted parties to catch colluders. In our context, sending fake collusion requests can reduce servers' confidence in others' compliance, but one needs to assume a trusted party to initiate dummy collusion. 

%% file: sections/10_conclusion.tex
\section{Concluding remarks and future directions}\label{sec:conclusion}
Following the intuition of a secret-shared bit guessing game, we design and implement a mechanism that discourages participants without network advantages from collusion to recover user secrets in multi-server PIR systems, which repeatedly provide query services to users, assuming a large $\ell$. The design itself rejects manipulation from users and servers with private knowledge. 

Because for practicality, we desire many servers to collectively provide the PIR service, blockchains and systems alike, where the available number of computing parties is much greater than what is needed for performing computations, become suited application scenarios. 

\paragraph{Future directions}
Our work opens up an interesting line of research. One intriguing direction for future work is to study mechanisms for other secret-sharing style applications or generic MPC. 
While the rationale behind our approach applies to many MPC tasks, the analysis demonstrates that many nuanced issues will have to be addressed both in the analysis and implementation of the evidence verification. 

%% file: sections/12_ack.tex
\section*{Acknowledgements}
We thank Vassilis Zikas for valuable feedbacks on system specifications, and anonymous reviewers and our shepherd for constructive comments. This work is supported partially by the National Science Foundation 
under grant CNS-1846316 and CAREER award CCF-2144208, and NSERC through the Discovery Grants program. 
Alexandros Psomas is also supported in part by a Google Research Scholar Award, a Google AI for Social Good award, and the Algorand Centres of Excellence program managed by the Algorand Foundation. Any opinions, findings, and conclusions or recommendations expressed in this material are those of the author(s) and do not necessarily reflect the views of Algorand Foundation.

%% file: sections/appendix-dishonest.tex
\section{Blind collusion problem}\label{sec:deceitful}

\paragraph{Forging information in collusion} When colluding parties do not verify the data shared by colluding partners, one interesting scenario is where a server prefers to retrieve the index alone or desires to hide its input to avoid being reported. 
We are interested in the \textit{probability of a colluding server fabricating information and fooling others}, i.e., other colluders mistake the provided fake data as correct, which we denote as  $p^\dagger$ hereafter. 

A cheating player guesses an index and tries to redirect the collusion protocol to output a targeted index. If the underlying PIR protocol ensures that fabricating information (a query or response) in collusion is computationally hard, then $\mathsf{p}_f$, the probability of creating a valid fake input, is 0. This would mean $p^\dagger=0$, and we are back to the discussion we have in the main body. 
We consider the worst case $\mathsf{p}_f=1$. This means that as long as a server can picture a redirecting strategy, it can always fabricate such an input.

\paragraph{Analysis}
We first discuss $k=2$. 
Let server $S_i$ receive query $q_1$ and $S_j$ ($i\neq j$) receive query $q_2$ such that $\D_a$ is the queried entry. To form a redirecting strategy, a server first guesses the intended query. One can guess $\D_a$ correctly with probability $\vec{x}_a$ where $\vec{x}_a$ is the probability that $\D_a$ is queried. Then the server creates a fake input to redirect the output to another index. We build the intuition with the naive xor-based PIR. We let $q_1=(j_1,j_2, \ldots ,j_m)$, $q_2 = (j_1,j_2,\ldots , j_m, a)$, and summarize the strategies in \Cref{tab:deceitfulplay}. 
\begin{table}[ht]
 \centering
 \caption[Redirecting strategy summary]{Redirecting strategy summary.}\label{tab:deceitfulplay}
 \scalebox{0.9}{
 \begin{tabular}{lll}
 \toprule
 \textbf{Strategy} & $S_i$ & $S_{j}$ \\ 
 \midrule
 Amicable & $j_1,j_2,\ldots,j_m$ &$j_1,j_2,\ldots,j_m,a$ \\
 $j_m\rightarrow a$ & $j_1,j_2,\ldots,j_{m-1},{\color{blue} a}$ & $j_1,j_2,\ldots,j_{m-1}$ \\
 $j_m\rightarrow j_{m+1}$ & $j_1,j_2,\ldots,j_{m-1},{\color{blue} j_{m+1}}$ & $j_1,j_2,\ldots,j_{m-1},a, {\color{blue} j_{m+1}}$ \\
 $j_m\rightarrow j_1$ & $j_2,\ldots,j_{m-1}$ & $j_2,\ldots,j_{m-1},a$ \\
 $a \rightarrow j_m$ & $j_1,j_2,\ldots,j_{m-1}, {\color{blue} a}$ & $j_1,j_2,\ldots,j_{m-1}$ \\
 $a\rightarrow j_{m+1}$ & $j_1,j_2,\ldots,j_m, {\color{blue} a}, {\color{blue} j_{m+1}}$ & $j_1,j_2,\ldots,j_{m},{\color{blue} j_{m+1}}$ \\
 $j_{m+1}\rightarrow j_{m+2}$& $j_1,\ldots,j_m, {\color{blue} j_{m+1}}, {\color{blue} j_{m+2}}$ & $j_1,\ldots,j_{m},a,{\color{blue} j_{m+1}}, {\color{blue} j_{m+2}}$ \\
 $j_{m+1}\rightarrow j_{m}$ & $j_1,\ldots,j_{m-1}, {\color{blue} j_{m+1}}$ & $j_1,\ldots,j_{m-1},a, {\color{blue} j_{m+1}}$ \\
 $j_{m+1}\rightarrow a$ & $j_1,\ldots,j_m,a, {\color{blue} j_{m+1}}$ & $j_1,\ldots,j_{m}, {\color{blue} j_{m+1}}$ \\ \bottomrule
 \end{tabular}
 }
\end{table}

Intuitively, if the outputs after the collusion protocol contain more than one index or gibberish when a server itself provides correct inputs, this server knows that the colluding party is cheating. 
Otherwise, if the collusion output contains only one index, the server may believe that collusion is successful. To find out this probability, we summarize four scenarios where a single index is outputted in \Cref{tab:deceitfuloutput}.

\begin{table}[ht]
 \centering
 \caption[Two cheating server collusion with single-index output]{Two-server collusion with single-index output.}
 \label{tab:deceitfuloutput}
 \begin{tabular}{ccc}
 \toprule
 Strategies
 & Possible single-index output & Probability $p^\dagger$ \\ \midrule
 $(A,A)$ & $a$ & 1 \\
 $(A,D)$ & $\{j_m, j_{m+1}\}$ & $\vec{x}_a(2-\vec{x}_a)$ \\
 $(D,A)$ & $\{j_m, j_{m+1}\}$ & $\vec{x}_a(2-\vec{x}_a)$ \\
 $(D,D)$ & $\{j_m, j_{m+1},j_{m+2},a,j_1\}$ & \begin{tabular}[c]{@{}l@{}}$\sum_i\sum_{j\neq i}4 \vec{x}_i^2 \vec{x}_j^2+$\\ $ \sum_{i\neq a}\sum_{j\neq i,a}4\vec{x}_a \vec{x}_i^2 \vec{x}_j$\end{tabular} \\
 \bottomrule
 \end{tabular}
\vspace{-2mm}
\end{table}

Row 1 is straightforward: when both servers input the correct information, index $a$ is retrieved with probability 1. In one cheating player case in rows 2 and 3, a single index is outputted when the cheating party either guesses the index correctly (with probability $\vec{x}_a$) or guesses the index incorrectly (with probability $(1-\vec{x}_a)$) but redirects the result to $\D_a$ (with probability $\vec{x}_a$). And $\vec{x}_a+(1-\vec{x}_a)\vec{x}_a=\vec{x}_a(2-\vec{x}_a)$, which we will denote as $p_{1}^\dagger$. Index $a$ is never output because the only cheating player either flips $a$ or produces more than one index in the output. The cheating player can always recover $a$ through deduction. 

In the case of two cheating players, by observing \Cref{tab:deceitfulplay}, we notice that when they have the same strategy or opposite strategies, they recover $a$. Because they cancel out each other's actions. This happens with probability $p_{2,1}^\dagger = \sum_i\sum_{j\neq i} 4 \vec{x}_i^2 \vec{x}_j^2$. For other cases of receiving a single output, $j_y\neq a$ becomes the output if one applies strategy $j_x \rightarrow a$ or $a \rightarrow j_x$ and the other cheating player plays a strategy $j_y \rightarrow j_x$ or $j_x \rightarrow j_y$. 
This occurs with probability $p_{2,2}^\dagger =\sum_{i\neq a}\sum_{j\neq i,a}4\vec{x}_a \vec{x}_i^2 \vec{x}_j$. 
To sum up, index $a$ is returned with probability $p_{2,1}^\dagger$ while a single index is outputted with probability $p_{2}^\dagger =p_{2,1}^\dagger +p_{2,2}^\dagger $.

Now we can extend our discussion to a more general $k>2$ case. When all servers play amicable, $a$ is output with probability 1. When one server cheats, the collusion protocol outputs a single index with probability $\vec{x}_a(2-\vec{x}_a)$. Similarly, index $a$ is never directly returned, and the only cheating player can always make successful accusations through deductions. 
When two servers cheat, we have the same probability calculation as before. 
When $2< k'\leq k$ servers cheat, they have a single output when each index appears in an even number of strategies, i.e., they have the same or opposite strategies, which occurs with probability $\sum_{i_1}\sum_{i_2\neq i_1}\cdots \sum_{i_{k'}\neq i_{k'-1}} 2^{k'} \vec{x}_{i_1}^2 \vec{x}_{i_2}^2\cdots \vec{x}_{i_{k'}}^2$ 
or when $a$ and exactly one other index $b\neq a$ appear in an odd number of strategies while every other index appears in an even number of strategies, which we compute next. Let $n_a, n_b$ be the number of appearances of $a$ and $b$ in strategies. We know that $n_a = 1, 3, \ldots, 2\lceil \frac{k'}{2}\rceil -1$, $n_b = 1, 3, \ldots, 2\lceil \frac{2 k' - n_a}{2}\rceil -1$ and $n_b\leq k'$. Let $n_c = 2k' - n_a - n_b$. 
For each $k'$, we have 
{\small 
\begin{multline*}
 p_{k'}^\dagger = \sum_{i_1}\sum_{i_2\neq i_1}\cdots \sum_{i_{k'}\neq i_{k'-1}} 2^{k'} \vec{x}_{i_1}^2 \vec{x}_{i_2}^2\cdots \vec{x}_{i_{k'}}^2 + \sum_{n_a} 2^{k'} \vec{x}_a^{n_a}\cdot \\
  \sum_{n_b} \vec{x}_{b}^{n_b} \sum_{i_1\neq a,b}\sum_{i_2\neq i_1,a,b}\cdots\sum_{i_{n_c}\neq i_{n_c-1},a,b} \vec{x}_{i_1}^2 \vec{x}_{i_2}^2\cdots \vec{x}_{i_{n_c}}^2
\end{multline*}
} 
In these scenarios with two or more cheating players, cheating servers cannot distinguish between different scenarios with a single output. 

\begin{table}[ht]
 \centering
 \caption[$k$-server collusion with single index output]{$k$-server collusion with single-index output. 
 }\label{tab:deceitfuloutput2}
 \scalebox{0.9}{
 \begin{tabular}{ccc}
 \toprule
 \#$D$ & Probability $p^\dagger$ \\ \midrule
 0 & 1 \\
 1 & $\vec{x}_a(2-\vec{x}_a)$ \\
 2 & $\sum_i\sum_{j\neq i}4 \vec{x}_i^2 \vec{x}_j^2+ \sum_{i\neq a}\sum_{j\neq i,a}4\vec{x}_a \vec{x}_i^2 \vec{x}_j$ \\
 $2<k'\leq k$ & $p_{k'}^\dagger $ \\
 \bottomrule
 \end{tabular}
 }
 \vspace{-4mm}
\end{table}

\subsection{Implications for the sequential game}\label{sec:gamenoverification}
Now $\mathsf{G}$ is a game of incomplete information: in round three, an agent is not certain of which actions the other agent has taken. Backward induction no longer applies, and we solve for sequential equilibrium instead. 

\paragraph{Information sets}
An \emph{information set} for a player $S_i$ contains all decision nodes, at which player $S_i$ takes an action, indistinguishable by $S_i$. 
Many information sets in our game are trivial, i.e., containing a single node. For example, in round 1, both agents are aware of the other agent's interests in collusion. However, there are non-trivial information sets, e.g., the ones prefixed ``12'' in \Cref{fig:demo2servers}, where the two make simultaneous decisions. 

\paragraph{Solution concept}
We consider another refinement of Nash equilibrium for sequential games with incomplete information, namely \textit{Sequential Equilibrium} \cite{kreps1982sequential}. 
A sequential equilibrium not only prescribes a strategy for each player but also a \emph{belief} about other players' strategies. 
To formalize the notion of sequential rationality, we first formalize the notion of a \emph{belief}. 
\begin{definition} For each information set $I$, a belief assessment $b$ gives the conditional probability distribution $b(\cdot|I)$ over all decision nodes $v \in I$.
\end{definition}
For example, $b(D_{2}|I^A_{1})$ is the belief of player $S_1$ that $S_2$ takes action $D$ in round one, given that $S_1$ is in information set $I^A_{1}=\{(A, A),(A, D)\}$. 
A belief assessment expresses all such beliefs for all players at their information sets. 

\begin{definition}
 A strategy profile and belief assessment pair $(\vec{s},b)$ is a sequential equilibrium if $(\vec{s},b)$ is sequentially rational and $b$ is consistent with $\vec{s}$.
\end{definition}
Given $(\vec{s},b)$, $\vec{s}$ is called \emph{sequentially rational} if at each information set $I$, the player to take an action maximizes her expected payoff, given beliefs $b(\cdot|I)$ and that other players follow $\vec{s}$ in the continuation (remaining) game. 
Given a strategy profile $\vec{s}$, for any information set $I$ on the path of play of $\vec{s}$, $b(\cdot|I)$ are \emph{consistent} with $\vec{s}$ if and only if they are derived using Bayes' rule. For example, let player 1 be at an information set $I=\{x,y\}$. Suppose according to $\vec{s}$, $x,y$ are respectively reached with probability $0.2, 0.3$. Then a consistent belief for player 1 is that $b(x|I)=\frac{0.2}{0.2+0.3}=0.4$ and $b(y|I)=0.6$.

We address the information sets in rounds one and two of $\mathsf{G}$ as information set $1$, $2$. They are singletons. In round three, we call the information set where one thinks others have played in round two as $3$, the information set where an amicable player recognizes that others have played $D$ as $4$, and the information set where a deceitful player recognizes that others have played $D$ as $5$. We denote reporting ($R$) at information sets $3,4,5$ as $R, R_4, R_5$. 
We have the following main theorem for the equilibrium of the collusion game without input verification. In a single-shot game, same as before, we let $\amtt{r}=0$. By enforcing $\amtt{s}>0, \amtt{f}>0$, we discourage false accusations.  
\begin{restatable}[]{theorem}{kaclnovrfy}
 In $2$-party collusion game $\mathsf{G}$ without input verification in $(\ell,2,1)$-PIR with $\theta \ell$ malicious servers, with probability $1-2^{-\eta}$, the pair $(\vec{s}, b)$ of strategy profile $\vec{s} = (\bar{C}DR\bar{R}_4 R_5, \ldots)$ for all servers and belief assessment $b$ is the sequential equilibrium if $\amtt{p} > 0$ 
 where $b$ satisfies $b( D_{\vec{s}_{-i}} | I^A_{3} ) = b( D_{\vec{s}_{-i}} | I^D_{3} ) = b( R_{\vec{s}_{-i}} | I^A_{3} ) = b( R_{\vec{s}_{-i}} | I^D_{3} ) = 1$, and $ \theta$ satisfies $S^\star$. 
 \label{prop: 2 agent l server case}
\end{restatable}

\begin{proof}[Proof]
 With probability $1-2^{-\eta}$, at least two queried servers are rational. Let the two queried servers be $S_1, S_2$. We depict the game tree in \Cref{fig:demo2serversnovrfy}. At information set 4 and 5, the dominant strategies are respectively $\bar{R}_4$ and $R_5$. 
 \begin{figure}[ht]
 \centering
 \includegraphics[width=0.45\textwidth]{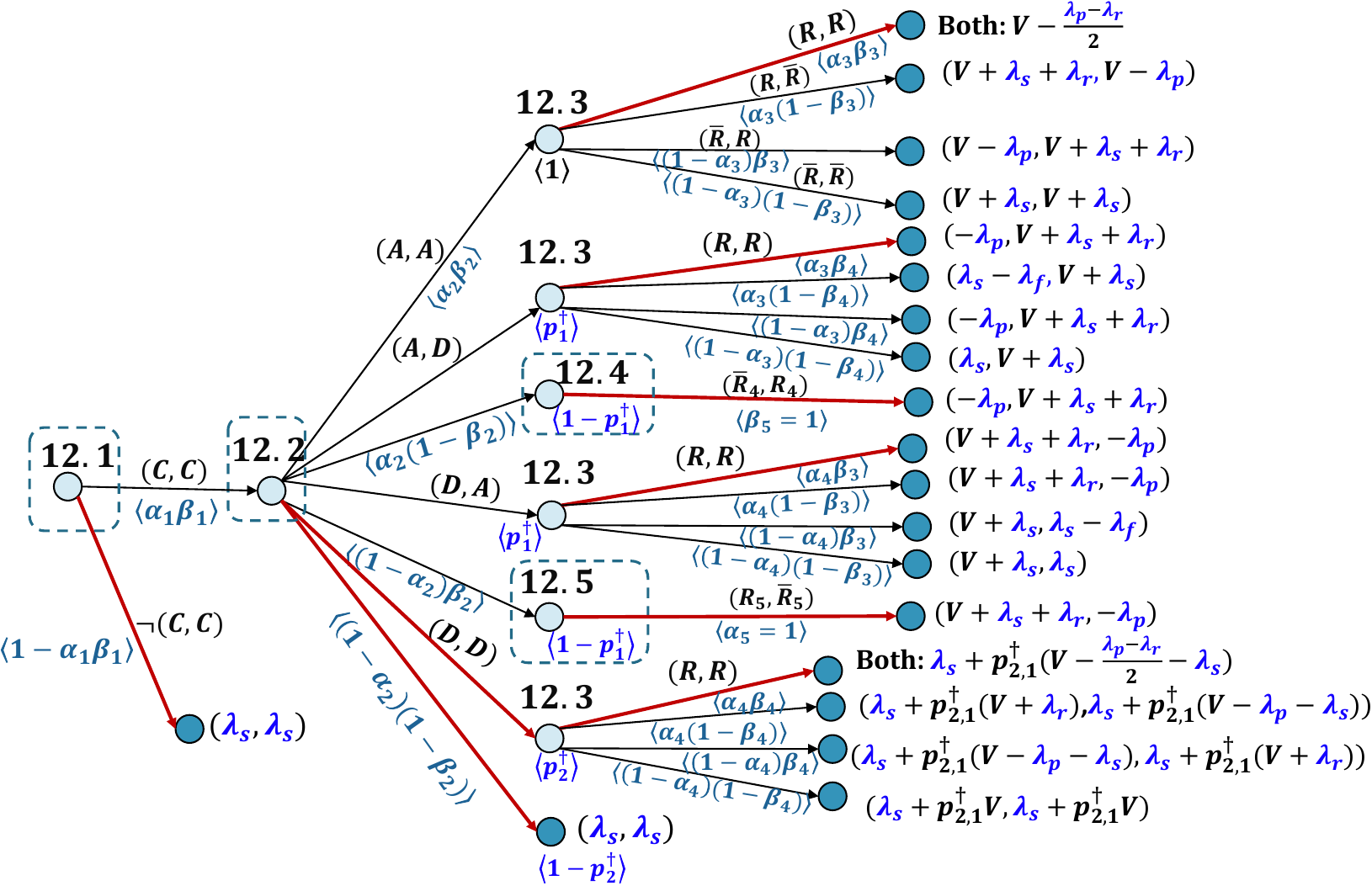}
 \caption{Extensive form of $\mathsf{G}$ without verification ($k=2$). 
 }
 \label{fig:demo2serversnovrfy}
 \end{figure}
 We then examine information set 3. It consists of four nodes. From top to bottom, the four nodes are reached with probability $\alpha_2 \beta_2, \alpha_2 (1- \beta_2) p_1^\dagger, (1- \alpha_2) \beta_2 p_1^\dagger, (1- \alpha_2) (1- \beta_2) p_2^\dagger$. Without loss of generality, we examine from one queried party $S_1$'s perspective. 
 By Bayes' formula, when $S_1$ plays $A$ at information set 2, it believes that it arrives at the information set 3 along move $(A, A)$ or $A_1 A_2$ with probability $b(A_2 | I_3^A) = \frac{ \beta_2}{ \beta_2 + (1 - \beta_2) p_1^\dagger}$. 
 Similarly, it believes it arrives at information set 3 along move $A_1 D_2$ with probability $b(D_2 | I_3^A) = \frac{ (1 - \beta_2) p_1^\dagger}{ \beta_2 + (1 - \beta_2) p_1^\dagger}$. When $S_1$ plays $D$ at information set 2, we can calculate $b(A_2 | I_3^D) = \frac{ \beta_2 p_1^\dagger}{ \beta_2 p_1^\dagger + (1 - \beta_2) p_2^\dagger}$ and $b(D_2 | I_3^D) = \frac{ (1 - \beta_2) p_2^\dagger}{ \beta_2 p_1^\dagger + (1 - \beta_2) p_2^\dagger}$. 
 For visual clarity, we denote the four probabilities as $x_1,x_2,y_1,y_2$. 
 If $S_1$ arrives at information set 3 after playing $D$, the expected returns are:
 {\footnotesize 
 \begin{multline*}
 y_1 \Big ( 
 \alpha_4 \beta_3 (\vcgmath + \amtt{s} + \amtt{r}) + \alpha_4 (1 - \beta_3) ( \vcgmath + \amtt{s} + \amtt{r}) +  (1 - \alpha_4) \beta_3 (\vcgmath + \amtt{s}) \\ +  (1 - \alpha_4) (1 - \beta_3) (\vcgmath + \amtt{s}) 
 \Big) + 
 y_2 \Big ( 
 \alpha_4 \beta_4 ( \amtt{s} + p_{2,1}^{\dagger} (\vcgmath - \frac{\amtt{p} - \amtt{r}}{2} - \amtt{s}) ) \\ + \alpha_4 (1 - \beta_4) \cdot  ( \amtt{s} + p_{2,1}^{\dagger} (\vcgmath + \amtt{r})) + (1 - \alpha_4) \beta_4 (\amtt{s} + p_{2,1}^{\dagger} (\vcgmath - \amtt{p} - \amtt{s})) \\ +  (1 - \alpha_4) (1 - \beta_4) (\amtt{s} + p_{2,1}^{\dagger} \vcgmath ) \Big) 
 \\ = \alpha_4 \big ( y_1 \amtt{r} + y_2 p_{2,1}^{\dagger} (\beta_4 (\amtt{p} + \amtt{r})/2 - \amtt{r} ) 
 \big ) + \epsilon = \alpha_4 y_2 p_{2,1}^{\dagger} \beta_4 \amtt{p}/2 + \epsilon 
 \end{multline*}
 }
Here $\epsilon$ is a term that contains only $\beta$s and does not depend on $S_1$'s strategies. 
Since $\amtt{p} > 0$, $\alpha_4=1$ maximizes the expected return. Similarly, $\beta_4 = 1$. 
Now we solve for $\alpha_3$ and $\beta_3$. Consider $S_1$ arriving at the information set 3 after playing $A$. There are three possible supports to consider for $S_1$: $\{R\}, \{\bar{R}\}, \{R, \bar{R}\}$ (randomization over $R,\bar{R}$). To make $R$ sequentially rational, we need the expected returns from playing $R$ to be higher than playing $\bar{R}$:
{\footnotesize 
 \begin{align*}
 &R: \; x_1 \big ( \beta_3 (\vcgmath - \frac{\amtt{p} - \amtt{r}}{2}) + (1-\beta_3)
 (\vcgmath + \amtt{s} + \amtt{r} ) \big ) - x_2 \amtt{p} \\
 &\bar{R}: \; x_1 \big ( \beta_3 (\vcgmath - \amtt{p}) + (1-\beta_3) (\vcgmath + \amtt{s}) \big ) - x_2 \amtt{p}
 \end{align*}
}
Then playing $R$ generates higher returns than $\bar{R}$, and $\alpha_3=1$. Similarly, $\beta_3=1$. 
We next solve for $\alpha_2$ and $\beta_2$. At information set 2, there are three possible supports to consider for $S_1$: $\{A\}, \{D\}, \{A, D\}$. To make $D$ sequentially rational, then we need the expected returns from playing $D$ to be higher than playing $A$. The simplified expected returns from playing the two moves are as follows:
{\footnotesize 
\begin{align*}
A: &\beta_2 (\vcgmath - \frac{\amtt{p} - \amtt{r} }{2}) + (1-\beta_2) \big ( p_1^\dagger ( - \amtt{p} ) + (1 - p_1^\dagger) (-\amtt{p}) \big ) \\
D: &\beta_2 (\vcgmath + \amtt{s} + \amtt{r}) + (1-\beta_2) \big ( p_2^\dagger p_{2,1}^\dagger (\vcgmath - \frac{\amtt{p} - \amtt{r} }{2} - \amtt{s}) + \amtt{s} \big )
\end{align*}
}
The first quantity is smaller than the second quantity since both terms are smaller. Because $\amtt{p}>0$ and $p_2^\dagger p_{2,1}^\dagger V + (1-\frac{p_2^\dagger p_{2,1}^\dagger}{2})\amtt{p} + (1-p_2^\dagger p_{2,1}^\dagger)\amtt{s}>0$. This means that $\alpha_2 = 0$. Likewise, $\beta_2=0$. Then $x_1=y_1=0, x_2=y_2=1$, and the above is consistent with the beliefs. 
In round one, playing $\bar{C}$ generates the same returns as $CD$. 
Then the sequential equilibrium is all players playing $\bar{C}DR\bar{R}_4 R_5$. 
\end{proof}

%% file: sections/appendix-constructions.tex
\section{Multi-Server PIR Constructions}\label{sec:constructions}

\paragraph{B.1. Boyle-Gilboa-Ishai's construction for $2$-server cPIR}
This construction utilizes a PRG-based 2-party Distributed Point Function (DPF). 
The client distributes the point function $f_{a,1}:[N]\rightarrow \mathbb{Z}_2$ to two servers to retrieve $\D_a$. She generates a pair of keys $(k_1,k_2)$ with a secure PRG, obtains function shares $(f_1,f_2)$ and sends $(f_i,k_i)$ to server $i$. Server $i$ responds with $\sum_{j=1}^N \D_jf_i(j)$. The client then obtains $\D_a$ by xor-ing two responses. 

\paragraph{B.2. $k$-server PIR}
We now work in an arbitrary finite field $\mathbb{F}$ and represent each entry in $\mathbb{F}$. Suppose it takes $e$ elements to describe each entry. Then $\D$ can be represented in $\mathbb{F}$ as an $e \times N$ matrix. We denote each entry as a vector $\vec{\D_i}$ in this section instead of $\D_i$ to signal this $e$-element representation. $\D_{ij}$ locates the $j$-th element of $\Vec{\D_i}$. 

\paragraph{Hafiz-Henry's construction for $k$-server cPIR} We present the protocol for querying $k=2^K$ ($K\in \mathbb{Z}^+$) servers.
The client intends to fetch $\D_a$ and needs to distribute the point function $f_{a,1}:[N]\rightarrow \mathbb{Z}_2$ to $k$ servers. 
She generates $K$ independent (2,2)-DPF key pairs, $(\mathsf{k}_0^{(0)}, \mathsf{k}_1^{(0)}), ..., (\mathsf{k}_0^{(K-1)}, \mathsf{k}_1^{(K-1)})$, for the point function $f_{a,1}$. 
The client now sends to each server $j$ the key share $(\mathsf{k}_{j_{K-1}}^{(K-1)}, ...,\mathsf{k}_{j_0}^{(0)})$. Each server $j$ expands each of these keys into a length-$N$ vector of bits and then concatenates the $K$ vectors component-wise to obtain a length-$N$ vector $\Vec{v}$ of $K$-bit strings. Server $j$ then goes through $\Vec{v}$ component by component, xor-ing the $\Vec{v}[i]$th word of $\D_i$ into a running total when $\Vec{v}[i]=1$. 
By construction, the $a$-th bit produced by each DPF key pair differs and all others are equal; thus, the XOR that each server produces is identical up to but not including which word of $\D_a$ it includes; moreover, one server includes no word of $\D_a$ at all. Taking this latter server’s response and xor-ing it with the responses from each of the other servers yields each of the words comprising $\D_a$ (in some random, but known to the querier, order).

For performance evaluation in \cite{hafiz2019bit}, the experiments are conducted on a workstation running Ubuntu 18.04.2 LTS on an AMD Ryzen 7 2700x eight-core CPU @ 4.30 GHz with 16 GiB of RAM and a 1 TB NVMe M.2 SSD. 
According to the results therein, sampling DPF keys on the clients' side takes less than 100 microseconds, and key expansion on the server side takes about 0.5ms for a database with $2^{20}$ rows. 
The answer reconstruction time is about constant for each extra bit retrieved. To reconstruct an entry of 1 GiB, the reconstruction time is about 130 ms. 

\paragraph{Efficiency comparison} SealPIR~\cite{angel2018pir} has $O(\sqrt{N})$ communication complexity, and incurs $\approx 100$x response size overhead and heavy computation (e.g., 400s for a 30 KB entry with $N=2^{20}$); OnionPIR~\cite{mughees2021onionpir} has $O(\log N)$ communication costs and results in $4.2$x response size overhead and similar computation costs; The Corrigan-Gibbss et al. proposal\cite{corrigan2022single} has $O(\sqrt{N})$ communication costs and achieves amortized sublinear server-side computation time by having the client download a one-time hint of size sublinear in $|\D|$; SimplePIR~\cite{henzinger2022one} achieves 6.5 GiB/s/core throughput but requires $O(\sqrt{N})$ communication costs, e.g., 124 MB hint plus 242 KB per query communication for a 1 GiB database. 

%% file: sections/appendix-selfinsurance.tex
\section{Self-Insurance}\label{sec: insurance}
\paragraph{A single server $S_e$ is exiting} 
Suppose users arrive at rate $\frac{\Omega}{T}$: during each time unit, there are $k \Omega$ queries. $S_e$ expects to receive $\frac{k \Omega}{\ell}$ queries and can recover all $\frac{k \Omega}{\ell}$ secrets in the worst case. The amount of accumulated service fees in those instances are $\sum_{t=1}^T k \frac{k \Omega}{\ell} \amtt{s} (1+r)^{T-t}$ in expectation. The expected amount of fines needed is $\sum_{t=1}^T \frac{k-1}{k} \frac{k \Omega}{\ell} \amtt{p} (1-r')^{T-t}$. 
We continue to let the ratio between inexecutable fines and service fees, $\frac{\sum_{t=0}^T \frac{k-1}{k} \frac{k \Omega}{\ell}\amtt{p} (1-r')^{T-t} - \amtt{p}}{\sum_{t=1}^T \frac{k \Omega}{\ell} \amtt{s} (1+r)^{T-t}} $, be $\leq 1$.

\paragraph{Multiple servers are exiting} 
Suppose $m \geq 2$ servers are leaving the service, and users arrive at rate $\frac{\Omega}{T}$. Let $b = \min \{m, k\}$, $i$ servers ($i = b, b-1, \ldots, 1$) are queried at the same time with probability $\frac{\binom{m}{i}\binom{\ell - m}{k - i}}{\binom{\ell}{k}}$. Expected inexecutable fines amount to $\sum_{t=0}^T \sum_{i=1}^{b} \frac{\binom{m}{i}\binom{\ell - m}{k - i}}{\binom{\ell}{k}} \frac{k-i}{k} \Omega i \amtt{p} (1-r')^{T-t} - m \amtt{p}$ and expected accumulated service fees are $\sum_{t=0}^T \sum_{i=1}^{b} \frac{\binom{m}{i}\binom{\ell - m}{k - i}}{\binom{\ell}{k}} \Omega k \amtt{s} (1+r)^{T-t}$. We then need $\frac{\sum_{t=0}^T \sum_{i=1}^{b} \frac{\binom{m}{i}\binom{\ell - m}{k - i}}{\binom{\ell}{k}} \frac{k-i}{k} \Omega i \amtt{p} (1-r')^{T-t} - m \amtt{p}}{\sum_{t=0}^T \sum_{i=1}^{b} \frac{\binom{m}{i}\binom{\ell - m}{k - i}}{\binom{\ell}{k}} \Omega k \amtt{s} (1+r)^{T-t}} \leq 1$.